\documentclass[journal]{IEEEtran}
\usepackage{cite}

\ifCLASSINFOpdf 
\usepackage[pdftex]{graphicx}
\else
\usepackage[dvips]{graphicx}
\fi

\usepackage[cmex10]{amsmath}
\usepackage{algorithmic}
\usepackage[ruled,norelsize]{algorithm2e}

\usepackage{array}

\usepackage{mdwmath}
\usepackage{mdwtab}
\usepackage{todonotes}

\usepackage{eqparbox}

\ifCLASSOPTIONcompsoc
\usepackage[caption=false,font=normalsize,labelfont=sf,textfont=sf]{subfig}
\else
\usepackage[caption=false,font=footnotesize]{subfig}
\fi

\usepackage{fixltx2e}
\usepackage{url}
\usepackage{graphicx}
\usepackage{amsfonts}
\usepackage{amssymb}
\usepackage{amsthm}
\usepackage{bm}
\usepackage[ruled]{algorithm2e}
\usepackage[utf8]{inputenc}
\usepackage{booktabs}
\usepackage{epstopdf}
\usepackage{tabularx}
\usepackage{cite}
\usepackage{flushend}

\definecolor{maroon}{cmyk}{0,0.87,0.68,0.32}

\newcounter{tempEquationCounter} 
\newcounter{thisEquationNumber}

\begin{document}
%


\title{Closing the Loop: A High-Performance Connectivity Solution for Realizing Wireless Closed-Loop Control in Industrial IoT Applications}


\author{
		Adnan~Aijaz,~\IEEEmembership{Senior~Member,~IEEE},
		and~Aleksandar~Stanoev
        \vspace{-0.5cm}
\thanks{The authors are with Toshiba Europe Ltd., Bristol, BS1 4ND, U.K. Contact e-mail: adnan.aijaz@toshiba-bril.com

Part of this work has been presented at the 27th International Conference on Real-Time Networks and Systems (RTNS), 2019 \cite{gallop_conf}. 
} 
}
\markboth{IEEE Internet of Things Journal -- Accepted for Publication}%
{Shell \MakeLowercase{\textit{et al.}}: Bare Demo of IEEEtran.cls for Journals}
%


\maketitle
\begin{abstract}
\boldmath
High-performance real-time wireless connectivity is at the heart of the Industrial Internet-of-Things (IIoT). Realizing wireless closed-loop control is crucial for various mission-critical IIoT systems.  Existing  wireless technologies fall short of meeting the stringent performance requirements of closed-loop control. This paper presents a novel wireless solution, called \textsf{GALLOP}, for  realizing closed-loop control over multi-hop or single-hop networks. \textsf{GALLOP} adopts a pragmatic design approach for addressing the challenges of wireless closed-loop control. Key design aspects of \textsf{GALLOP} include control-aware bi-directional scheduling for cyclic exchange, robust retransmission techniques based on cooperative multi-user diversity and low-overhead signaling for scalable operation. \textsf{GALLOP} has been specifically designed for control loops that are closed over
the whole network with dynamics on the order of few
milliseconds. Performance evaluation based on extensive system-level simulations and hardware implementation on a Bluetooth 5 testbed demonstrates that \textsf{GALLOP} provides high-performance connectivity with very  low and deterministic latency, very high reliability and high scalability, to meet the stringent requirements of  wireless closed-loop control for versatile IIoT applications. 

\end{abstract}



%
\IEEEpeerreviewmaketitle

\section{Introduction}

\IEEEPARstart{T}{he} recent \emph{Industrial} Internet-of-Things (IIoT) \cite{IoT_SLR}  initiative is broadly focusing on application of connectivity technologies in industrial domains such as manufacturing, transportation, logistics, and energy.  Unlike consumer IoT systems, IIoT systems focus on connecting industrial assets like machines and control systems with information technologies and business processes \cite{IIoT_TII, ind_internet}. IIoT systems are mission-critical in nature and require high-performance connectivity to fulfil the stringent requirements of timeliness and reliability \cite{hrllc}. 

Closed-loop control or feedback control is a prominent industrial control application. Typically, it takes place between a controller and a spatially distributed system of sensors and actuators. Control and feedback signals are exchanged in the form of information packets over a shared wireless medium, thereby closing a global control loop. In legacy industrial systems, closed-loop control is realized through wired technologies like  Industrial Ethernet \cite{ind_Ethernet}. However, such technologies have high installation and maintenance costs. Multi-hop wireless communication based on low-power wireless devices provides a cost-effective and flexible connectivity solution. However, state-of-the-art multi-hop wireless technologies like WirelessHART, ZigBee and ISA 100.11a do not fulfil the challenging connectivity requirements of industrial closed-loop control applications. Such technologies are mainly used for non-critical applications like monitoring of large-scale industrial systems. 

Designing a multi-hop wireless network for closed-loop control becomes particularly challenging due to its stringent connectivity requirements. Closed-loop control has strong real-time requirements and demands connectivity with very high reliability and very low latency \cite{TI_PIEEE}. This is because packet losses and delays, which are dominant in wireless environments, have a detrimental effect on the stability of control loops. Closed-loop control involves bi-directional communication with cyclic traffic patterns. The cyclic information exchange demands highly deterministic connectivity which implies that the communication latency (between cycles) must have a very low variance. The presence of a large number of sensors and actuators also creates the requirement of highly scalable connectivity.

Against this background, the objective of this paper is to design a high-performance connectivity solution for realizing wireless closed-loop control over multi-hop networks. Our focus is specifically on \emph{wireless design for control loops that are closed over the entire multi-hop network with dynamics on the order of few milliseconds (ms)}. We propose a novel wireless solution, termed as \textsf{GALLOP} (\underline{G}ener\underline{A}lized c\underline{L}osed-\underline{L}oop c\underline{O}ntrol of \underline{P}rocesses) that has been designed to address the challenges of multi-hop closed-loop control. Although  \textsf{GALLOP} has been designed for multi-hop networks, it is fully capable of operating in single-hop scenarios.  \textsf{GALLOP} design encompasses various novel features (Section \ref{sect_ka}), which provide high-performance wireless connectivity with very low and deterministic latency, very high reliability and high scalability. \textsf{GALLOP} can be extended to a broad range of IIoT applications. 
The key contributions of this work are summarized as follows.   

\begin{itemize}

\item  \textsf{GALLOP} handles cyclic information exchange through multi-hop time division multiple access (TDMA) scheduling. We design a novel \emph{distributed}  multi-hop scheduling algorithm for \textsf{GALLOP} which is described in Section \ref{sect_dpo}. 


\item We design novel techniques for handling medium access control (MAC) layer transmission failures in \textsf{GALLOP}. These techniques, which are discussed in Section \ref{sect_ret}, ensure reliable operation in harsh wireless environments. 

\item We develop an analytic framework (Section \ref{sect_ana}) for characterizing  convergence aspects of distributed scheduling. 

\item We conduct a comprehensive evaluation of \textsf{GALLOP} through system-level simulations (Section \ref{sect_perf}). 

\item We provide experimental results (Section \ref{sect_imp}) for \textsf{GALLOP}  based on hardware implementation on a state-of-the-art Bluetooth 5 platform and evaluation on a real Bluetooth 5 testbed. We also benchmark the performance against a standard solution. 

\end{itemize}

\textcolor{black}{An early version of this work appeared in \cite{gallop_conf}. This paper provides a number of new contributions including extensions to protocol design, analysis of convergence time and further evaluation based on system-level simulations and hardware implementation.  }



\section{Related Work}
The problem of wireless closed-loop control can be approached from two different perspectives: \emph{control-aware wireless design} and \emph{wireless-aware control design}.  
The control-aware wireless design paradigm, which is the focus of this work,  aims to design network protocols for meeting the stringent connectivity requirements of control-centric applications. Existing solutions for control-aware wireless are mainly focused on single-hop connectivity. Wireless Interface for Sensors and Actuators (WISA) \cite{wisa} is one of the earliest solutions designed for industrial automation networks. The Wireless Sensor Actuator Network for Factory Automation (WSAN-FA) standard \cite{wsan_fa} adopted WISA with certain modifications. WISA as well as WSAN-FA target a latency  of \(~10\) ms for up to \(120\) devices. However, both solutions offer insufficient reliability for the target latency performance. \textcolor{black}{Both WISA and WSAN-FA have also influenced the development of the recent IO-Link Wireless standard \cite{IOLW} which targets a latency of \(5\) ms for up to \(40\) devices in single-hop. }
Through novel enhancements and optimizations, \textsf{ENCLOSE} \cite{enc_TII} outperforms WISA, WSAN-FA and other similar industrial wireless solutions in terms of reliability, latency, scalability, and coverage.    Control-centric wireless solutions based IEEE 802.11 standard (Wi-Fi) have also been developed. Prominent solutions include Industrial Wireless Local Area Network (IWLAN) \cite{S_IWLAN} and Wireless Network for Industrial Automation - Factory Automation (WIA-FA) \cite{IEC_WIA-FA}. The latest IEEE 802.11ax standard is particularly attractive for industrial wireless applications as it provides much lower latency compared to previous Wi-Fi generations \cite{aijaz_IES}. EchoRing \cite{echoring} achieves highly reliable and responsive connectivity based on a token-passing algorithm. Even though it extends beyond single-hop connectivity, it is not as scalable as WISA and WSAN-FA. The 5G mobile/cellular technology is also a candidate for control-centric industrial applications \cite{5G_FA, pvt_5G}. 

On the contrary, the objective of the wireless-aware control design paradigm is to develop control algorithms to cope with wireless imperfections.
Alur \emph{et al.} \cite{mh_cont} developed a compositional framework for modeling and analysis of multi-hop control networks. 
The framework separates control, topology, routing, and scheduling aspects and allows co-design of control algorithms and communication parameters. 
Smarra \emph{et al.} \cite{opt_codesign} addressed the problem of guaranteed stability with co-design of a digital controller and network parameters.
Pajic \emph{et al.} \cite{wi_cont_net} proposed a new approach that treats the multi-hop network as a controller. 
 Tatikonda and Mitter \cite{cont_comm_tac} investigated bounds on  required channel
rates to achieve different control objectives. 
Robinson and Kumar \cite{loc_cont} investigated the problem of minimizing packet drops through optimization of controller location. Zhang \emph{et al.} \cite{survey_net_ind} surveyed different control methodologies for tackling network-induced imperfections. 


Multi-hop TDMA scheduling in wireless networks has been extensively surveyed in a recent study \cite{survey_TDMA}.  Multi-hop scheduling in \textsf{GALLOP} is unique in various aspects. First, it has been designed for closed-loop control that entails bi-directional exchange, which has been rarely considered in prior art, particularly in the case of distributed techniques. Second, distributed scheduling in \textsf{GALLOP} adopts a pragmatic approach and implements a novel signaling mechanism. Existing scheduling techniques
seldom consider such practical aspects. Third, scheduling  in \textsf{GALLOP} is particularly optimized to minimize cyclic  latency. Fourth, scheduling in \textsf{GALLOP} provides native support for  cooperative transmissions that are crucial for high reliability. Last, but not least, scheduling in \textsf{GALLOP} handles MAC layer transmission failures while accounting for the peculiarities of closed-loop control.

\section{\textsf{GALLOP} -- Design and Protocol Operation} \label{sect_dpo}
\subsection{Key Aspects} \label{sect_ka}
\textsf{GALLOP} has been designed for high-performance wireless connectivity in order to realize closed-loop control over multi-hop networks. The centerpiece of \textsf{GALLOP} design is a \emph{control-aware bi-directional schedule} that handles  cyclic exchange of control informaton between a controller and a spatially distributed system of sensors
and actuators. To reduce cycle time, the scheduling algorithm exploits both parallel and sequential transmissions for minimizing the cycle time  which is a key metric for closed-loop control. \textsf{GALLOP} builds a distributed schedule through a \emph{low-overhead signaling} approach that exploits local overhearing. To achieve fast convergence, signaling is performed jointly for both downlink and uplink. \textsf{GALLOP} implements \emph{robust and efficient retransmission techniques} that ensure high reliability. Some of its retransmission techniques harness \emph{cooperative multi-user diversity} by integrating cooperative and multi-user diversity techniques. The retransmission techniques also exploit  opportunities for network coding that provides low latency and high reliability. \textsf{GALLOP}   also supports \emph{frequency/channel hopping} for mitigating the impact of external interference. The design of \textsf{GALLOP} is agnostic to the Physical (PHY) layer. Our implementation is based the on the Bluetooth PHY layers. However, it can be implemented on any other PHY
layer (IEEE 802.15.4, IEEE 802.11, etc.).

\subsection{Scenario, Assumptions and PHY/MAC Layer Design}
\begin{figure}
\centering
\includegraphics[scale=0.32]{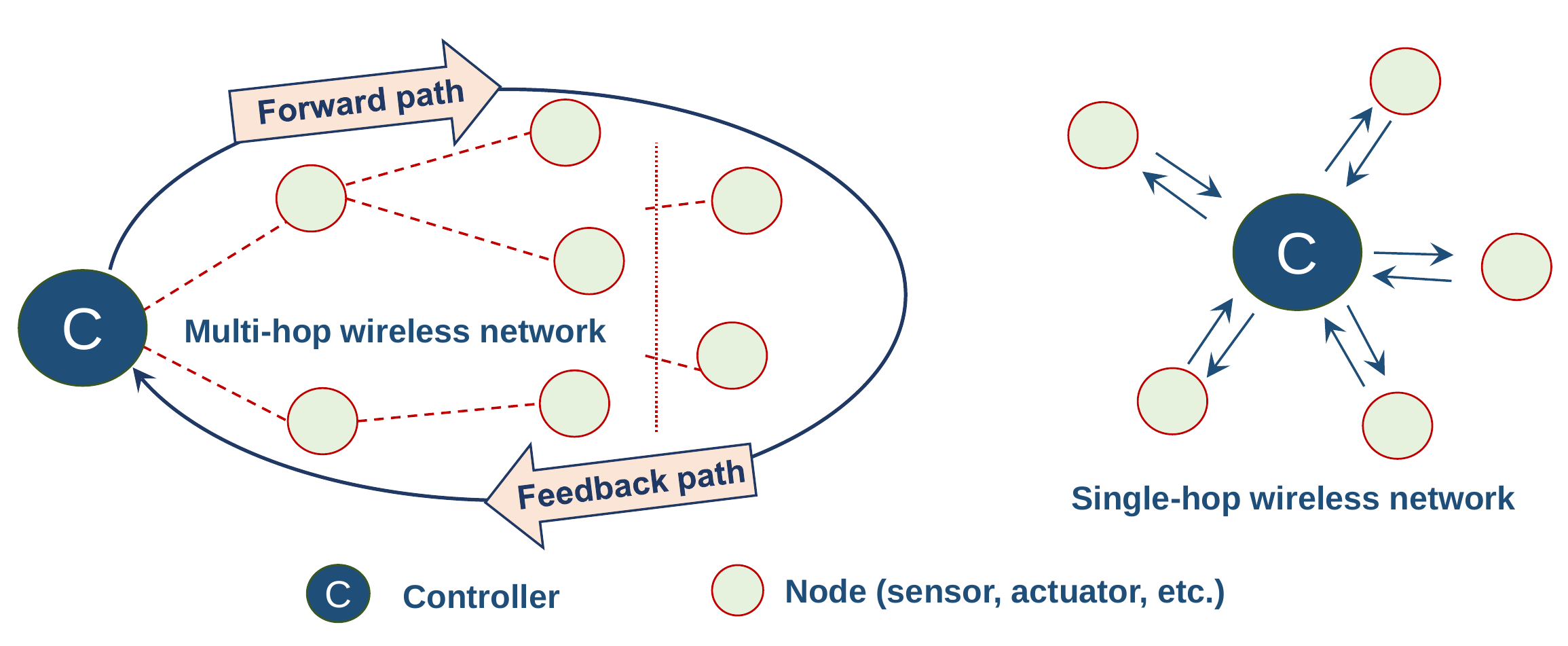}
\caption{\textcolor{black}{Multi-hop and single-hop control scenarios for \textsf{GALLOP}.}  }
\vspace{-0.5cm}
\label{top}
\end{figure}

We consider a closed-loop control scenario which  is shown in Fig. \ref{top}. \textcolor{black}{Multiple devices (sensors, actuators, etc.) exchange bi-directional and cyclic information with a controller over a multi-hop or a single-hop wireless network.} This scenario reflects a broad range of mission-critical IIoT applications. Some of the key applications are described as follows.
\begin{itemize}
\item In industrial automation, cyclic exchange of command and feedback messages occurs between a  controller and multiple robotic arms. 
\item In case of vehicle platooning, a lead vehicle (controller) maintains constant velocity/distance for the followers  through closed-loop interaction. Similar principle can be applied to leader-follower formation control \cite{comms_formation} of automated guided vehicles in logistics and warehouse operations.
\item In a large-scale battery network, a battery management unit (controller) monitors and controls individual battery cells through cyclic exchange of command and feedback messages. 
\end{itemize}

For protocol operation, the terms \emph{downlink} and \emph{uplink} refer to communication from the controller  to nodes/devices and vice versa, respectively.  Without any network layer protocol, the nodes constitute a mesh topology. 
We assume that the mesh topology has been converted into a tree topology through any low-cost routing protocol (e.g., RPL \cite{RPL}).  The tree topology provides multi-hop connectivity between the controller and the devices. We consider a simple interference model where two nodes are one-hop neighbors if their mutual distance is less than the communication or interference range. For achieving network-wide time synchronization, \textsf{GALLOP} utilizes a flooding-based protocol, known as Glossy \cite{glossy}. The flooding process is initiated by the controller and driven by radio events. It exploits synchronous/concurrent transmissions and entails relaying overheard packets immediately upon reception. Time synchronization based on Glossy has been described in detail in \cite{glossy}.	It is emphasized that other time synchronization techniques can also be adopted for \textsf{GALLOP} implementation. 

%

Our implementation of \textsf{GALLOP}  is based on the PHY layer of Bluetooth technology which  is quite promising for industrial communication \cite{TI_PIEEE}. Although \textsf{GALLOP} can be realized on both basic rate (BR) and low energy (LE) Bluetooth radios, our  
implementation is on the latter. 
The LE radio provides a data rate of \(1\) Mbps (referred to as LE 1M PHY). The latest Bluetooth 5 standard introduces two new PHY layers: LE 2M PHY which supports a data rate of \(2\) Mbps, and LE Coded PHY that supports a data rate of \(500\)kbps or \(125\) kbps. 
The MAC layer of \textsf{GALLOP} is based on the principles of multi-hop TDMA and  frequency division duplexing (FDD). There are three different MAC frames: a downlink frame for downlink communication, an uplink frame for uplink communication and 
a signaling frame for exchange of schedule-related
information.
 Each frame comprises a certain number of timeslots of fixed duration. 


\begin{figure}
\centering
\includegraphics[scale=0.265]{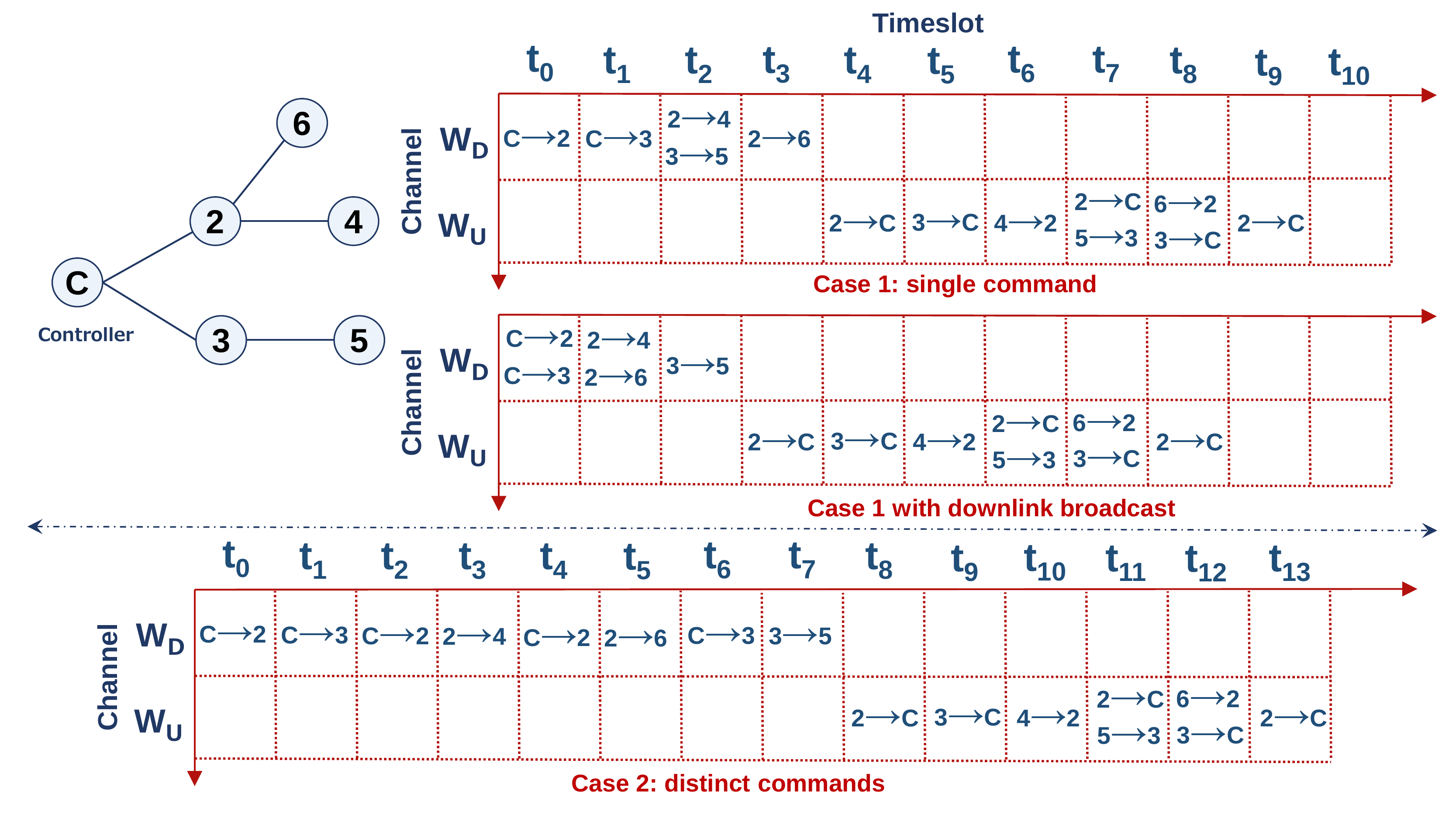}
\caption{Centralized bi-directional multi-hop scheduling in \textsf{GALLOP}. }
\vspace{-0.5cm}
\label{cent_sch}
\end{figure}


\subsection{Centralized Bi-directional Multi-hop Scheduling}
To elaborate on the scheduling problem for multi-hop wireless closed-loop control, we discuss  a centralized solution where the controller  is responsible for deriving a schedule based on global knowledge of network topology. 
 We consider the topology  in Fig. \ref{cent_sch} where the connectivity (neighborhood) information at the MAC layer for each node is  as follows. 
\begin{equation}
\begin{cases}
C: [2, 3], \  \text{node 2}: [C, 4, 6], \ \text{node 3}: [C, 5], \\
\text{node 4}: [2, 6], \ \text{node 5}: [3], \  \text{node 6}: [2, 4]
\end{cases}
\label{top_ni}
\end{equation}


The  controller's objective is to minimize \emph{cycle time} which is defined as the cumulative time  of successfully transmitting command messages to all the nodes in the network and successfully receiving the  response message from each node. We consider the case where the controller transmits a single command message for the network.  The controller implements a simple scheduling algorithm where downlink and uplink transmissions are scheduled in a sequential manner while exploiting parallel transmission opportunities. 
The constructed schedule is shown in Fig. \ref{cent_sch}. First, the controller sequentially schedules 
downlink transmissions (command messages) on a downlink channel \(W_D\). The downlink transmissions from nodes 2 and 3 are non-conflicting; hence, these can go in parallel. After downlink scheduling, which takes \(4\) timeslots, the controller initiates uplink scheduling on the uplink channel \(W_U\).  Non-conflicting uplink transmissions are scheduled in parallel. The overall cycle time  is \(10\) timeslots. Note that the controller can also schedule downlink transmissions in a broadcast manner as illustrated in Fig. \ref{cent_sch}. This decreases the cycle time by a timeslot. 
 The scheduling for distinct command messages (also shown in Fig. \ref{cent_sch}) for each node follows a similar approach with the exception of  scheduling distinct downlink transmissions. 


%
%
%
%

Due to the overhead of acquiring topological information and schedule dissemination, centralized scheduling techniques offer limited scalability. \textcolor{black}{Therefore, the main focus of this work is distributed scheduling. }


\subsection{Distributed Bi-directional Multi-hop Scheduling}
\begin{figure}
\centering
\includegraphics[scale=0.29]{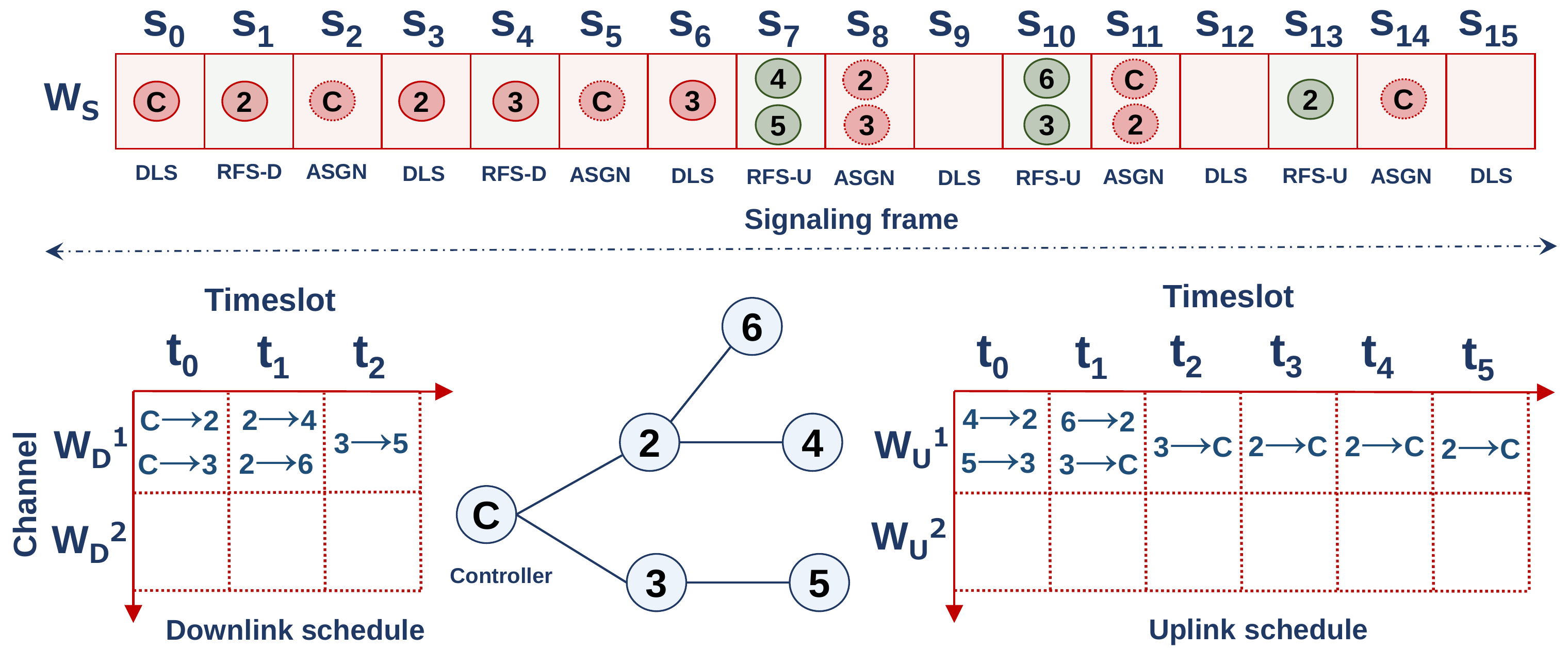}
\caption{\textcolor{black}{Distributed bi-directional multi-hop scheduling in \textsf{GALLOP}.} }
\vspace{-0.5cm}
\label{dist_sch1}
\end{figure}

In the distributed case, there is neither a central entity  nor global knowledge of network topology. Schedule synthesis occurs via  neighbor-to-neighbor information exchange and negotiation. The fundamental principle of \textsf{GALLOP}'s distributed scheduling is to construct a conflict-free schedule through parent-child handshake and local overhearing. Such information exchange occurs over a signaling frame  (signaling channel). \textsf{GALLOP} implements different types of signaling messages. The \emph{request-for-slots} (RFS) message is used by a child node to request resources (e.g., timeslots) from its parent node. Different RFS messages are employed in downlink and uplink, i.e., RFS-Downlink (RFS-D) and RFS-Uplink (RFS-U), respectively. In response to the RFS message, a parent node uses an  \emph{assignment} (ASGN) message  to allocate resources to a child node. The \emph{downlink signaling} (DLS) message is used by a parent node to disseminate schedule-related information to its child nodes. The signaling frame contains three different types of signaling slots. The first slot (\(s_0\)) is a DLS slot and reserved for the controller. This is followed by a repetitive pattern (illustrated in Fig. \ref{dist_sch1}) of three slots such that the RFS slots, the ASGN slots and the DLS slots are indexed by \(\left\lbrace s_1, s_4, s_7, \cdots \right\rbrace\), \(\left\lbrace s_2, s_5, s_8, \cdots \right\rbrace\) and \(\left\lbrace s_3, s_6, s_9, \cdots \right\rbrace\), respectively. 

\begin{figure}
\centering
\includegraphics[scale=0.41]{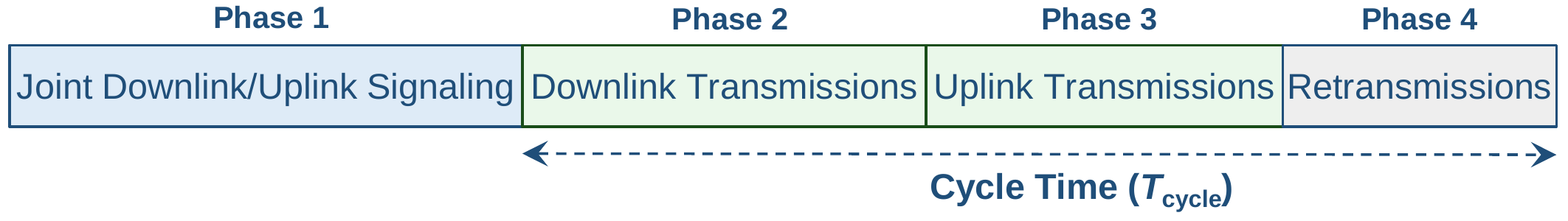}
\caption{\textcolor{black}{Different phases of \textsf{GALLOP} protocol operation.} }
\vspace{-0.5cm}
\label{phases}
\end{figure}

Fig. \ref{phases} shows different phases of \textsf{GALLOP} operation. The first phase is joint downlink/uplink signaling where schedules for downlink and uplink transmissions are built.  
An alternate approach is to decouple the signaling such that it occurs separately for downlink and uplink schedules. However,
our experience of hardware implementation (discussed later) reveals that joint downlink/uplink signaling not only decreases the signaling phase duration but also reduces the cycle time.

Let \(\mathcal{M}=\left\lbrace 1, 2, \cdots, M \right\rbrace\) denote the set of nodes of the multi-hop network. The set of parent nodes and leaf nodes that constitute the network is denoted by \(\mathcal{R}\) and \(\mathcal{F}\), respectively.  A parent node has at least one child node associated with it at the network layer. The leaf nodes in the network are those with no child nodes. We assume that each parent node knows the number of its child nodes based on the information from the network layer. 

\textsf{GALLOP} builds a sequential schedule in order to minimize the latency for cyclic information exchange. This implies that, in the downlink, the schedule begins from the controller and evolves towards the leaf nodes. Similarly, in the uplink it begins from the leaf nodes and evolves towards the controller. Sequential scheduling becomes feasible due to the unique signaling mechanism of \textsf{GALLOP}. 
Nodes maintain \emph{local knowledge} of available timeslots and channels through overhearing of different signaling messages during the signaling phase. The distributed scheduling process exploits local knowledge and follows a set of policies and rules that dictate schedule construction.

\textbf{Policy for Downlink Signaling} -- This policy dictates access to the signaling channel and plays an important role in overall schedule construction. It consists of a single rule. 

\textbf{\emph{Rule DL-1:}} \emph{Any parent node \(r \in \mathcal{R}\)  assigns priorities to its child nodes  and conveys the information regarding next available RFS slot using the DLS message. }


The priority  of a node can be based on any unique parameter such as its rank at the network-layer or its MAC identifier. The prioritization of child nodes in any order does not affect the schedule building process. The child nodes use the priority information for conflict-free access to the signaling frame. 

\textbf{Policy for Parent-Child Handshake} -- The objective of this policy is conflict-free resource allocation in the multi-hop network. It is based on three different rules. 

\textbf{\emph{Rule H-1:}} \emph{Any child node \(m  \in \mathcal{M}\) requests the earliest possible timeslots on a given channel through the RFS message where it finds no conflict as per its local knowledge. }

\textbf{\emph{Rule H-2:}} \emph{Upon reception of an RFS message, any parent node \(r \in \mathcal{R}\)  confirms the requested allocation by transmitting an ASGN message if it finds no conflict based on its local knowledge. }

\textbf{\emph{Rule H-3:}} \emph{If a parent node \(r \in \mathcal{R}\) detects a conflict in requested allocation, it responds with an updated allocation, in the ASGN message, as per its local knowledge of earliest available timeslots on a given channel. }

We explain the role of these two policies in building the downlink schedule through the topology shown in Fig. \ref{dist_sch1} that has the same MAC layer neighborhood/connectivity relationship as given by \eqref{top_ni}. 
The  signaling process is initiated by the controller through transmission of a DLS message in the first signaling slot (\(s_0\)) of the signaling channel \(W_S\). As per the policy for downlink signaling, it assigns priorities to its child nodes (nodes 2 and 3). We assume that a lower node ID\footnote{\textcolor{black}{A parent node knows the node IDs of its child nodes from the network setup phase when the routing tree is generated.}} corresponds to a higher priority. Therefore, node 2 has priority over node 3. The DLS message from the controller also contains the index of the next available RFS slot (denoted by \(ix\_RFS\_slot\)), i.e., \(s_1\). Any child node computes  the index of its RFS slot as 
\(ix\_RFS\_slot+priority-1\).
Since node 2 has the highest priority, it determines its RFS slot as \(s_1\). Node 3 determines its RFS slot as \(s_4\), as it has the second priority. Note that the controller is exempt from handshake. In the downlink, the controller (and any other parent node) can either schedule its child nodes in a unicast (distinct timeslots) or a broadcast (single timeslot) manner. We focus on the latter case as the former is straightforward extension of the proposed mechanism. For downlink transmission, the controller selects timeslot \(t_0\) on the downlink channel \(W_D\). The schedule information of the controller is also included in the DLS message. In case of distinct transmissions, the controller would have selected timeslots \(t_0\) and \(t_1\). As per node 2's local knowledge, timeslot \(t_0\) is not available. Hence, based on Rule H-1, it picks timeslot \(t_1\) and transmits an RFS-D message in slot \(s_1\). As  the controller does not detect a conflict, it confirms the allocation as per Rule H-2  by transmitting an ASGN message in the next ASGN slot \(s_2\). After successful downlink allocation, any parent node uses the first available DLS slot to transmit the DLS message as per the policy for downlink signaling. Hence, node 2 selects slot \(s_3\) to transmit a DLS message to its child nodes (nodes 4 and 6).  The DLS message also contains the index of the next available RFS slot. Node 2 has the information regarding the number of child nodes of the controller (based on the received DLS message). It determines that the first two RFS slots would be taken by the child nodes of the controller. Therefore, the next available RFS slot for its child nodes is \(s_7\). In slot \(s_4\), node 3 sends an RFS-D message to the controller. Nodes 2 and 3 are not in the neighborhood of each other. We assume that the ASGN message in slot \(s_2\) (from the controller) was not successfully overheard by node 3. Hence, as per Rule H-1, node 3 requests timeslot \(t_1\) on the downlink channel \(W_D\) from the controller. However, based on controller's local knowledge, this timeslot is taken by node 2. Therefore, as per Rule H-3, the controller allocates timeslot \(t_2\) through the ASGN message in slot \(s_5\). After successful downlink allocation, node 3 uses the next available DLS slot, i.e., slot \(s_6\) to transmit a DLS message to node 5. Based on the local knowledge of node 3, the next available RFS slot for node 5 is \(s_7\). This completes successful construction of the downlink schedule as all  parent nodes have timeslots allocated for downlink transmissions.  

In \textsf{GALLOP}, the uplink schedule construction process  is initiated by the leaf nodes. Owing to joint downlink/uplink signaling, leaf nodes trigger uplink signaling (for schedule construction) that follows a set of policies and rules.

\textbf{Policy for Uplink Signaling} -- This policy dictates access to the signaling channel for construction of the uplink schedule. It consists of the following rules. 


\textbf{\emph{Rule UL-1:}} \emph{Any  parent node \(r \in \mathcal{R} \ (r \neq \text{controller})\) requests uplink timeslots only if it has scheduled all of its child nodes in the uplink.}

\textbf{\emph{Rule UL-2:}} \emph{Any  parent node \(r \in \mathcal{R} \ (r \neq \text{controller})\) requests \(\mathcal{S}_r+1\) timeslots in the uplink, where \(\mathcal{S}_r\) is the total number of timeslots requested by its child nodes. }

Building a sequential schedule in uplink is a key aspect of the uplink signaling policy. Based on the DLS message from node 2, nodes 4 and 6 determine \(s_7\) and \(s_{10}\) as their RFS slots, respectively. Similarly, node 5 determines its RFS slot as \(s_7\) based on node 3's DLS message. Nodes 4 and 5 trigger the uplink schedule construction process in slot \(s_7\) as these are leaf nodes. As per Rule H-1, both nodes select timeslot \(t_0\) on the uplink channel \(W_U\) and transmit an RFS-U message to their  parent nodes. Note that neither node 2 can overhear node 5 nor node 3 can overhear node 4, as they are not in the neighborhood of each other. Therefore, both nodes 4 and 5 can share the respective RFS slot without causing collisions at the respective receivers. As per Rule H-2, nodes 2 and 3  confirm the requested allocation through ASGN messages in slot \(s_8\). Based on node 6's local knowledge, timeslot \(t_0\) is occupied. Hence, it selects timeslot \(t_1\) and transmits an RFS-U message in slot \(s_{10}\). As per Rule H-2, Node 2 confirms the requested allocation by transmitting an ASGN message in slot \(s_{11}\). 

After successful allocation to node 5 in slot \(s_8\), node 3 can now request timeslots for uplink communication as per Rule UL-1. Based on its local knowledge, the next available RFS slot is \(s_{10}\). As per Rule UL-2, it requests two timeslots in an RFS-U message, i.e., \(t_1\) and \(t_2\) on the uplink channel \(W_U\). The controller confirms the requested allocation via an ASGN message in slot \(s_{11}\). Node 2 is ready to request uplink timeslots after scheduling its child nodes. Based on its local knowledge, the next available RFS slot is \(s_{13}\). Note that node 2 is not aware of the allocation for node 3. Since it requires three timeslots (\(\mathcal{S}_2=2\)), it requests \(t_2\), \(t_3\) and \(t_4\) on the uplink channel \(W_U\)  via an RFS-U message. As the controller detects a conflict on timeslot \(t_2\) that has been allocated to node 3, it responds with an updated allocation, i.e., timeslots \(t_3\) -- \(t_5\) on the uplink channel \(W_U\), via an ASGN message in slot \(s_{14}\).  
This completes the uplink scheduling as all nodes have timeslots allocated for uplink transmissions.

\begin{figure}
\centering
\includegraphics[scale=0.265]{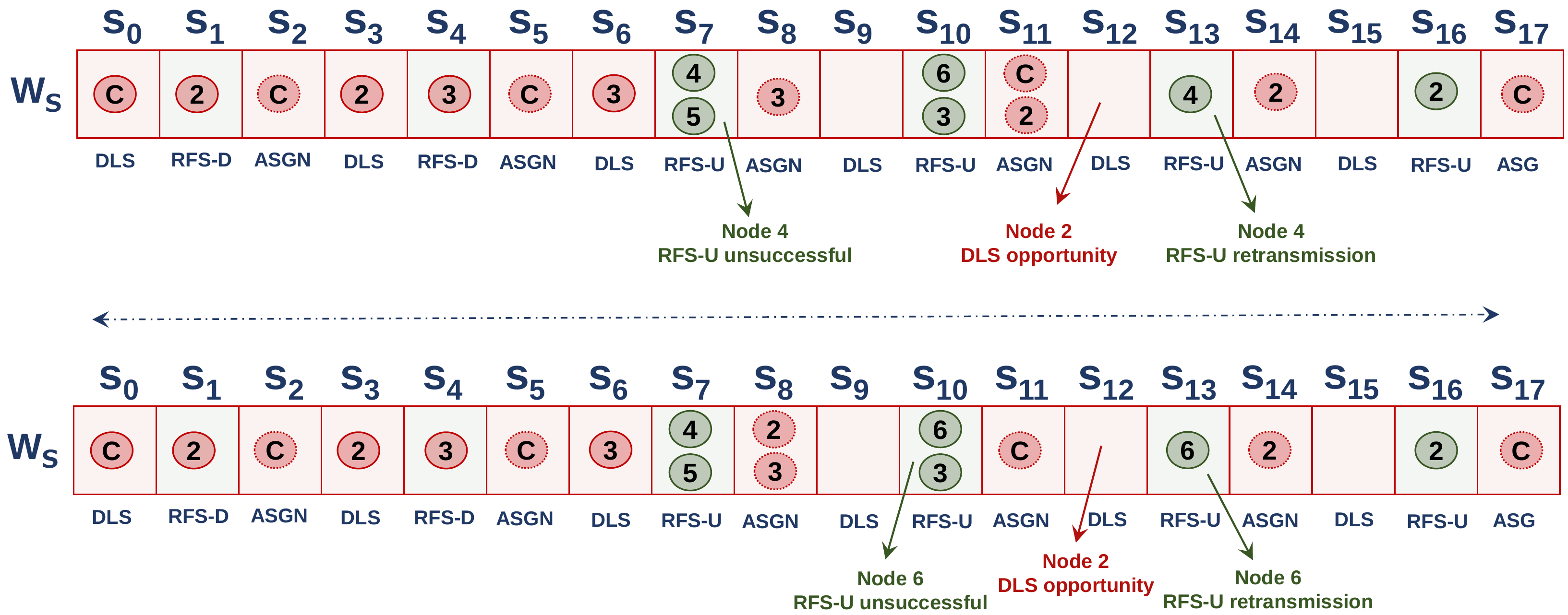}
\caption{\textcolor{black}{Illustration of the signaling message retransmission policy. The topology is same as in Fig. \ref{dist_sch1}.}   }
\vspace{-0.5cm}
\label{sig_retrans}
\end{figure}

\textbf{Policy for Signaling Message Retransmission} -- This policy dictates access to the signaling frame in case of signaling message retransmissions. During the schedule construction phase, RFS and DLS messages are retransmitted under the following conditions. 

\begin{itemize}
\item Any node \(m \in \mathcal{M} \ (m \neq \text{controller})\) retransmits an RFS message if it does not receive an ASGN message from its parent node. This happens due to a  collision in an RFS slot or a link-level RFS/ASGN transmission failure. 

\item Any parent node \(r \in \mathcal{R}\) retransmits a DLS message if it does not receive an RFS message from at least one of its child nodes within the  \emph{allocation window}, which is given by \(ix\_RFS\_slot+3 \times \Theta_r -1\), where \(\Theta_r\) is the number of child nodes and \(ix\_RFS\_slot\) is the index of the RFS slot as advertised in the original DLS message. 
\end{itemize}

The policy for signaling message retransmission consists of the following rules.

\textbf{\emph{Rule SR-1:}} \emph{Any  node \(m \in \mathcal{M} \ (m \neq \text{controller})\) retransmits the RFS message (on first attempt) in an RFS slot whose index is determined as \(ix\_RFS\_slot+\Theta_r+priority-1-U_r\), where  \(\Theta_r\) and \(U_r\) denote the number of child nodes and the number of successfully allocated child nodes of its parent node \(r\), respectively. 
}

\textbf{\emph{Rule SR-2:}} \emph{Any  node \(m \in \mathcal{M} \ (m \neq \text{controller})\) retransmits the RFS message (on second or higher attempt) in an RFS slot whose index is determined as \(ix\_prev\_RFS\_slot+B_m+priority-1\), where \(ix\_prev\_RFS\_slot\) is the index of the last used RFS slot and \(B_m\) denotes a small random backoff that is uniformly distributed in the interval \(\left[1, 2, \cdots, \Psi \right]\).
}

Rule SR-1 ensures that an RFS retransmission does not disrupt the handshake process of a sibling node, i.e., a node having the same parent. It also ensures that a node does not have to wait for an arbitrary long duration before retransmitting an RFS message by accounting for the number of successfully allocated sibling nodes. This information is acquired through retransmitted DLS messages. Moreover,
Rule SR-2 potentially avoids conflicts between multiple nodes in neighborhood of each other that require RFS retransmissions. 

\textbf{\emph{Rule SR-3:}} \emph{Any parent node \(r \in \mathcal{R}\) retransmits the DLS message if an opportunity for retransmission arises within the allocation window. Otherwise, on first  attempt, it retransmits the DLS message  in the first available DLS slot as per its local knowledge, and on second or higher attempt, in a DLS slot whose index is given by  \(ix\_prev\_DLS\_slot+B_m-1\), where \(ix\_prev\_DLS\_slot\) is the index of the last used DLS slot and \(B_m\) denotes a small random backoff that is uniformly distributed in the interval \(\left[1, 2, \cdots, \Psi \right]\).}

Note that the leaf nodes do not transmit DLS messages. Therefore, after a parent node has ascertained the need for DLS retransmission, a retransmission opportunity arises if any of its child nodes transmits an RFS-U message during the allocation window.  An opportunity  also arises if a parent node does not receive an RFS message from a child node in its assigned  slot. Rule SR-3 ensures that a child node awaiting DLS for requesting resources does not have to wait for an arbitrary long duration while mitigating contention between multiple parent nodes requiring DLS retransmissions in neighborhood. 
A parent node embeds information of the number of successfully allocated child nodes in the retransmitted DLS message. 
This  is used by a child node to compute appropriate RFS slot for retransmission as per Rule SR-1. 

The retransmission procedure is illustrated in Fig. \ref{sig_retrans}. Consider that the RFS-U message from node 4 is not successfully received by node 2 in slot \(s_7\). Node 4 is aware that node 2 has two child nodes. As per Rule SR-1, it retransmits the RFS-U message in slot \(s_{13}\). Fig. \ref{sig_retrans} also shows the scenario where RFS-U message from node 6 is not successfully received by node 2 in slot \(s_{10}\). Without  DLS retransmission from node 2 in slot \(s_{12}\), node 6 would have retransmitted the RFS-U message in slot \(s_{16}\). However, the DLS retransmission from node 2 (as per Rule SR-3), informs node 6 that node 4 has been successful. Hence, it retransmits the RFS-U message in slot \(s_{13}\), as per Rule SR-1. 


\textbf{Policy for Schedule Termination} -- The scheduling phase is terminated by the controller once all of its child nodes have been successfully scheduled in the uplink.

\begin{figure}
\centering
\includegraphics[scale=0.28]{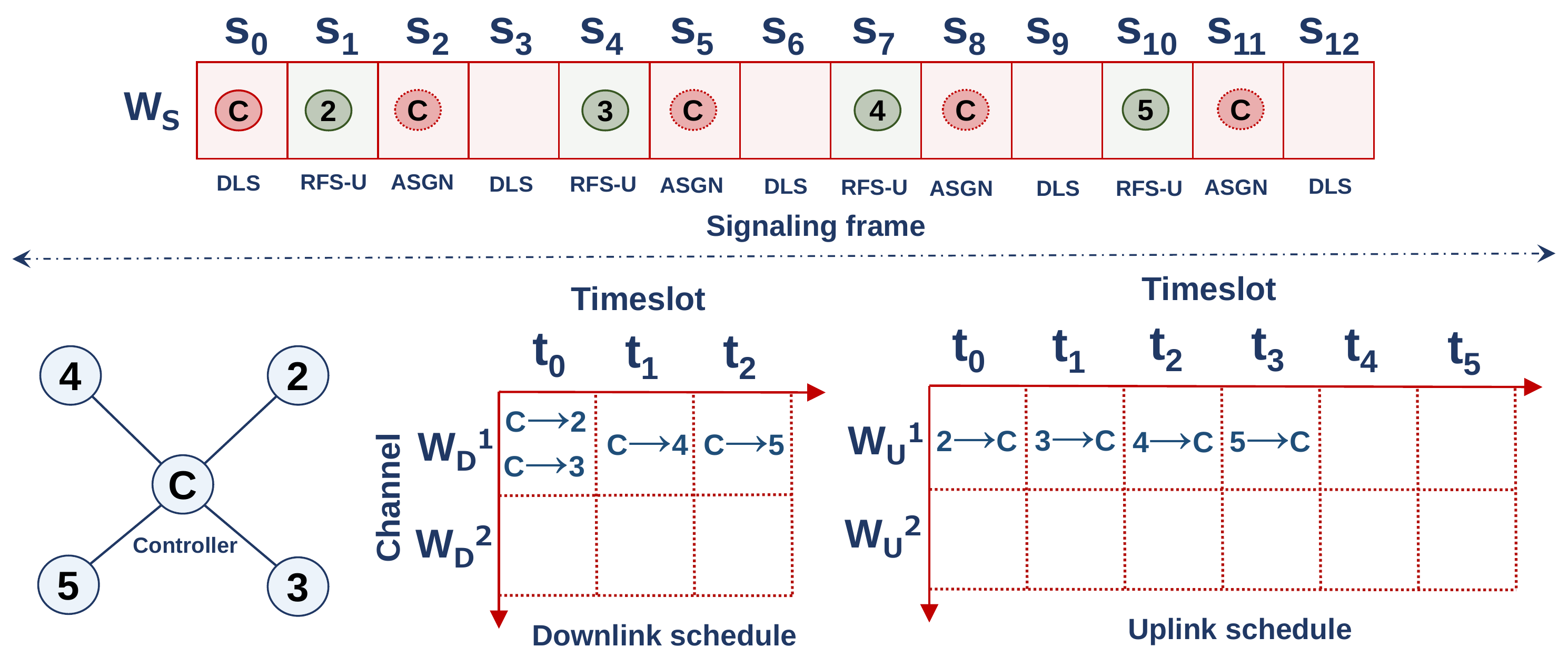}
\caption{\textcolor{black}{Distributed scheduling of \textsf{GALLOP} for single-hop operation. }  }
\vspace{-0.5cm}
\label{dist_sh}
\end{figure}

\subsection{\textcolor{black}{Single-hop Operation}}
\textcolor{black}{Although \textsf{GALLOP} has been designed for multi-hop operation, it also has the capability of operating in single-hop environments.  The proposed distributed scheduling framework is transparent to the network topology. Hence, it can be directly applied in single-hop scenarios without any modifications. Fig. \ref{dist_sh} illustrates the distributed scheduling procedure and the constructed schedule for single-hop operation.   The scheduling process, which is initiated by the controller, follows the aforementioned policies for downlink and uplink signaling and parent-child handshake. We assume that the controller has distinct messages for nodes 2 and 3, node 4 and node 5. Hence, it takes three timeslots on the downlink channel. In the uplink, each leaf node is allocated a distinct timeslot for communication with the controller.    }

\subsection{Frequency/Channel Hopping}
\textsf{GALLOP} implements two different frequency/channel hopping techniques. In \emph{phase-slotted} hopping, frequency hopping is performed on a phase-by-phase basis. The actual frequency/channel for the phase pertaining to data transmissions or retransmissions is given by \(\left\lbrace \left(AP\_No+C\_{Off}\right) \ \textrm{mod} \ N_C \right\rbrace\), where \(AP\_No\) is the absolute phase number, \(C\_{Off}\) is the channel offset, \(N_C\) is the total number of channels and \(\textrm{mod}(.)\) denotes the modulo operation. 
In \emph{time-slotted} hopping frequency/channel hopping is performed at timeslot level in a similar manner. 

\section{\textsf{GALLOP} -- Retransmission Techniques} \label{sect_ret}
In \textsf{GALLOP},  cyclic information exchange between the controller and  slaves/devices occurs over the built downlink and uplink schedules. A transmission in either downlink or uplink may fail due to interference or fading. Retransmissions, which are handled in the final phase of \textsf{GALLOP}, are crucial in achieving high reliability.  Note that recovery of MAC layer transmission failures in multi-hop control networks  lead to the overhead-determinism-reliability (ODR) trade-off.  \textsf{GALLOP} implements three different MAC layer retransmission techniques that distinctively account for the ODR trade-off. 

\subsection{Schedule Duplication}
\begin{figure}
\centering
\includegraphics[scale=0.28]{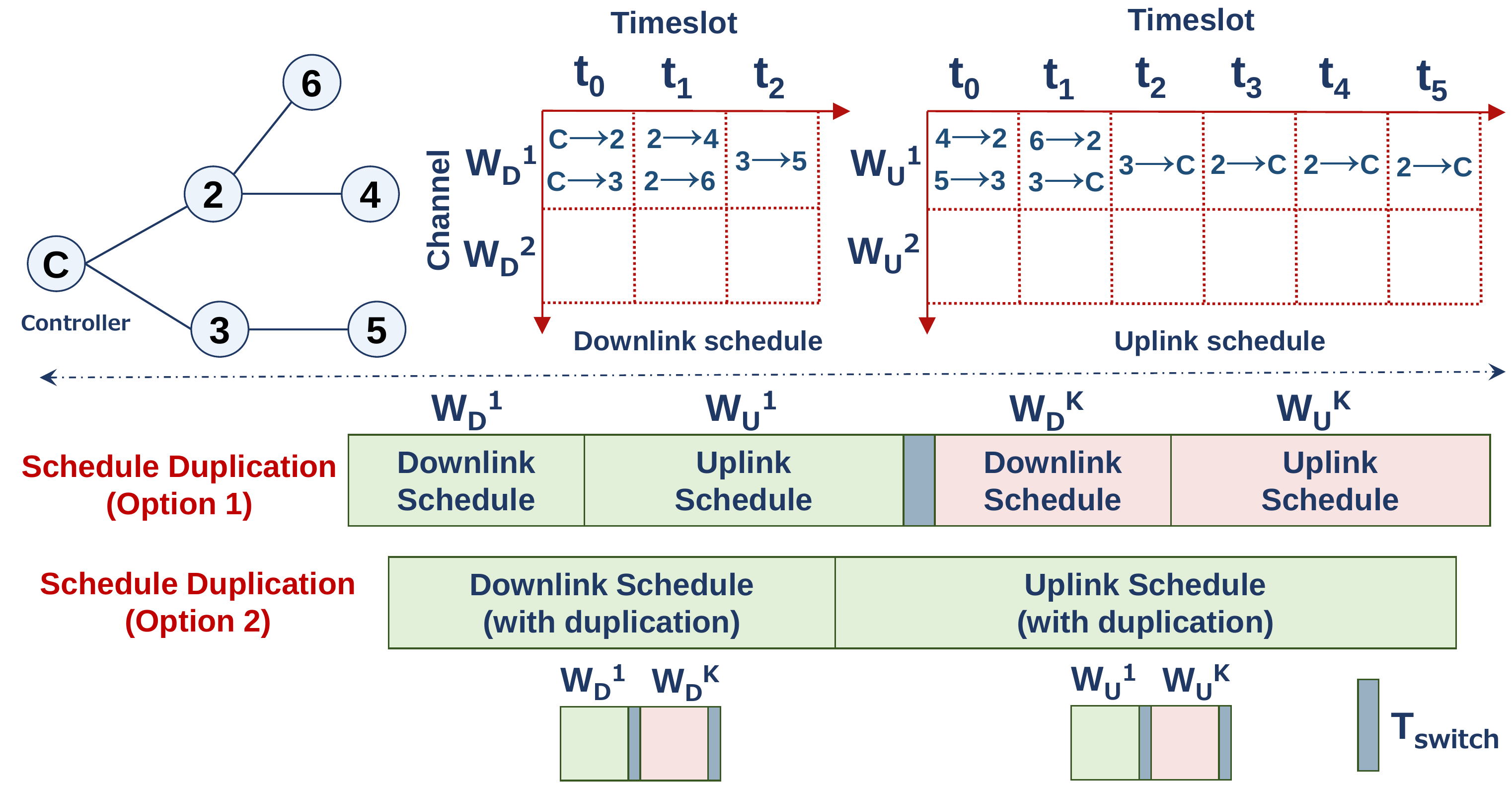}
\caption{\textcolor{black}{Schedule duplication technique and the two options for duplication.}  }
\vspace{-0.5cm}
\label{sch_dup}
\end{figure}

The first technique is a manifestation of packet duplication functionality, which  is achieved through duplicating the transmission schedule in time domain. Further, this technique integrates channel diversity with packet duplication. The two transmissions of the same packet at two different time instances and on two different frequencies provides high reliability. The main motivation for schedule duplication is to guarantee deterministic performance. Besides, it does not incur any overhead. Schedule duplication  can be achieved in two different ways which are illustrated in  Fig. \ref{sch_dup}.
The \emph{first option} is to  duplicate the transmission schedule. The first round of transmissions takes places as per the constructed schedule. In the second round, the downlink and uplink  transmission phases are repeated, as per the existing schedule, with each phase going on a different channel. Note that there is a fixed channel switching delay between individual rounds. While this approach improves reliability in a deterministic manner, it also doubles the latency. The schedule can be repeated more than once as long as the latency bound is not violated. The \emph{second option}  is to build the schedule in a way that accounts for packet duplication. During the schedule construction phase, each node requests one extra timeslot per transmission, which is used for duplication. To incorporate channel diversity, the second transmission of the same packet goes on a different channel. This requires a fixed channel switching delay between adjacent timeslots. This approach also provides higher reliability in a deterministic manner. However, its latency could be lower than the previous option due to the fact that \textsf{GALLOP} exploits opportunities for parallel transmissions in the schedule.


\subsection{Retransmission Scheduling}
The second  technique builds a schedule for  MAC layer transmission failures. It also exploits cooperative multi-user diversity and opportunistic network coding techniques. Note that in multi-hop wireless networks, any parent node needs to forward its own  information along with that of its child nodes. In \textsf{GALLOP}, 
a parent node maintains a certain order of transmissions on the allocated (uplink) timeslots.
It  uses the first timeslot for its own information. Subsequent timeslots are used for forwarding the  information of child nodes in the order of priority. For example, in  Fig. \ref{dist_sch1} node 2 uses timeslot \(t_3\) for transmitting its own  information, and  \(t_4\) and \(t_5\) for forwarding the information of nodes 4 and 6, respectively. Each parent node also enumerates packets in  sequence of transmission, e.g., node 2 will mark its transmissions in timeslots \(t_3\), \(t_4\), and \(t_5\) as \(1\), \(2\), and \(3\), respectively.
To exploit cooperative multi-user diversity, \textsf{GALLOP} relies on certain nodes to act as relays for others. The relay selection method is described later. The relay node for any arbitrary node is selected by its parent node from  its child nodes. The relay nodes overhear during the transmission schedule phase and buffer the overheard packets from the nodes for which they act as relays. The need for building a retransmission schedule is ascertained by the controller. At the end of the transmission schedule, the controller determines if a transmission from any of its child nodes is missing. The missing transmission could be of the controller’s child or of any other node further away in the multi-hop topology. If one or more transmissions are found missing, the controller generates a \textsf{GALLOP} negative acknowledgement (G-NACK) message which contains the IDs of its child nodes with failed transmissions along with the number of the missing transmission. The subsequent parent nodes generate their own G-NACK messages if required. The G-NACK messages are disseminated in the network  through the downlink schedule. 
The signaling for  retransmission schedule follows the same handshake policy as explained earlier. Unlike the original schedule building phase, there are only two types of signaling slots on the signaling channel, i.e., RFS slots and ASGN slots which are numbered as  \(\left\lbrace s_0, s_2, s_4, \cdots \right\rbrace\) and \(\left\lbrace s_1, s_3, s_5, \cdots \right\rbrace\), respectively. 

\textbf{Retransmission Scheduling Signaling Policy} -- The retransmission scheduling signaling policy dictates signaling channel access for building the retransmission schedule. It consists of the following rules.

\textbf{\emph{Rule RS-1:}} \emph{Any  node \(m \in \mathcal{M} \ (m \neq \text{controller})\) accesses the signaling channel in the earliest available slot, as per its local knowledge, if it has the missing transmission in its  buffer.}

\textbf{\emph{Rule RS-2:}} \emph{Any parent node \(r \in \mathcal{R} \ (r \neq \text{controller})\) with a failed transmission from any of its child nodes accesses the signaling channel in the earliest available slot, as per its local knowledge, after it has allocated timeslots for the missing transmissions.}


\textbf{\emph{Rule RS-3:}} \emph{The controller terminates the signaling after it has allocated timeslots for all the missing transmissions.}

The retransmission scheduling technique is  explained by distinguishing the following cases which are illustrated in Fig. \ref{ret_sch}. Note that the topology in Fig. \ref{ret_sch} is different than that in Fig. \ref{dist_sch1} as the controller has an additional child node (node 4 which only has the controller and node 2 in its  neighborhood). Hence, the constructed schedule is slightly different.

\subsubsection{Case A}
Assume that at the end of the transmission schedule, the second transmission from node 2 was missing at the controller. The controller sends a G-NACK message to its child nodes in the first timeslot of the downlink schedule. Based on the G-NACK message, node 3 determines that it does not need to engage in  retransmission scheduling phase. However, node 2 needs to participate in the retransmission scheduling  to recover its missing transmission. It knows that the second transmission (indicated in the G-NACK) was meant to be used for forwarding node 6’s information. This transmission could be in its own buffer, provided the  transmission from node 6 to node 2 in timeslot \(t_0\) of the uplink schedule  was successful. We assume that this transmission was successful and it is available in node 2's buffer. Therefore, it sends an empty G-NACK message to its child nodes in the second timeslot of the downlink schedule. The empty G-NACK notifies its child nodes that their transmissions were successful. Similarly, node 3 sends an empty G-NACK to node 5 in the third timeslot of the downlink schedule. After the end of the downlink schedule phase, all nodes switch to the signaling channel.
Based on Rule RS-1, node 2 will access slot \(s_0\) of the signaling channel \(W_S\) to request timeslots for the missing transmission through an \emph{RFS retransmission scheduling} (RFS-RS) message. To exploit cooperative multi-user diversity, each node with a failed transmission in its buffer will request two timeslots. Hence, node 2 requests timeslots \(t_0\) and \(t_1\) on the retransmission channel \(W_R\). Since the controller finds no conflict, it confirms the requested allocation as per Rule H-2 by sending an ASGN message in slot \(s_1\). After successful allocation, the controller terminates the scheduling phase, as per Rule RS-3, by sending a specific downlink message. 

After termination of the retransmission signaling phase, nodes switch to the retransmission channel. Recall that node 2 requested two timeslots. The first timeslot is used by a parent node to communicate with the relay. We assume that node 4 is the default relay  for node 2. As relay nodes overhear, it is likely that the missing transmission from node 2 is in node 4's buffer. The missing transmission can be recovered via various types of cooperative transmissions employing the relay node.

\begin{figure}
\centering
\includegraphics[scale=0.27]{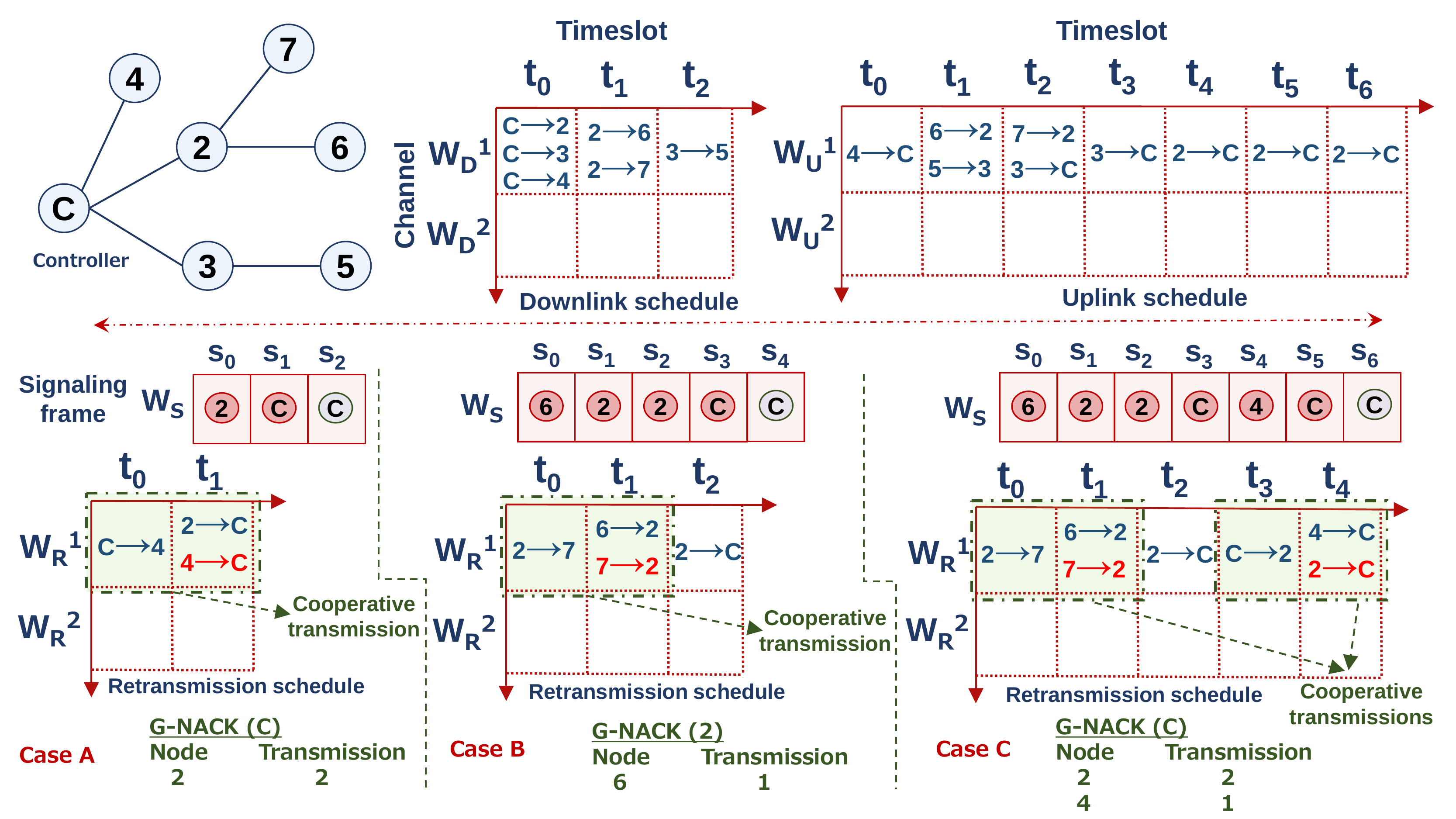}
\caption{\textcolor{black}{Retransmission scheduling technique and the schedule construction process in different cases.}  }
\vspace{-0.5cm}
\label{ret_sch}
\end{figure}

 \textbf{Cooperative Transmission 1} --  The source node and the relay node simultaneously transmit the missing information. For instance, as shown in Fig. \ref{ret_sch}, node 2 and node 4 simultaneously transmit in timeslot \(t_1\). This provides reliability enhancement by providing a gain in received power, if the two transmissions are tightly time synchronized within half the symbol period.  This phenomenon, which has been  exploited by a number of recent synchronous/concurrent transmission-based flooding protocols \cite{CI_tutorial,zimm_survey}, is referred to as \emph{constructive interference} (CI) in literature. 
 

\textbf{Cooperative Transmission 2} -- The relay node performs a simple network coding (SNC) operation by encoding the missing information with any \emph{a priori} information at the controller. Let, \(Y_{22}\) denote the missing transmission (second transmission of node 2). The relay node encodes \(Y_{22}\) and its own uplink transmission, \(Y_{41}\) (that was successfully received by the controller), by performing an XOR operation (i.e., \(Y_{22} \oplus Y_{41}\)), and transmitting the encoded information in timeslot \(t_1\) to the controller. From a reliability perspective, the SNC operation provides two distinct benefits: (i) simple XOR-based coding has been shown to introduce a gain as compared to the case of simply forwarding overheard information \cite{error_nc}, and (ii) relay's transmission  provides path diversity.

%


It is also possible to integrate cooperative transmissions 1 and 2.
Besides, cooperative multi-user diversity can  be achieved through  physical layer network coding (PLNC) \cite{PLNC_survey}. Similar to cooperative transmission 1, node 2 and node 4 simultaneously transmit \(X_{22}\) in timeslot \(t_1\); however, the controller decodes the missing information based on \emph{a priori} mapping for interpreting the combined received signal. 

\subsubsection{Case B}
Similar to the previous case, the second transmission from node 2 was missing at the controller;  however, it was not available in the buffer of node 2. Therefore, node 2 sends a G-NACK to its child nodes indicating that the first transmission from node 6 is missing. After the downlink schedule phase ends, nodes switch to the signaling channel. As node 6 is the leaf node, it is the source node for the missing transmission. In this case we assume that node 6 sent an uplink message (in timeslot \(t_1\)); however, it was not successfully received by node 2. As per Rule RS-1, node 6 requests timeslots \(t_0\) and \(t_1\), on the retransmission channel \(W_R\), by sending an RFS-RS message in slot \(s_0\). Node 2 confirms the requested allocation in slot \(s_1\), as per Rule H-2. As the transmission from node 6 needs to be forwarded to the controller, node 2 requests timeslot \(t_2\) by sending an RFS-RS message in slot \(s_2\), as per Rule RS-2. After confirming the requested allocation, the controller terminates the scheduling phase. 
We assume that node 7 is the default relay for node 6.  Therefore, timeslots \(t_0\) and \(t_1\) are used for cooperate transmission, exploiting CI, to recover the failed transmission of node 6. In the subsequent timeslot (\(t_2\)), node 2 forwards this information to the controller. Node 2 can either simply transmit this information or it can perform a SNC operation.

\subsubsection{Case C}
Consider that at the end of the transmission schedule, the second transmission from node 2 and the only transmission from node 4 were missing at the controller. The controller indicates these  in the G-NACK message to its child node. Similar to the previous case, the missing transmission from node 2 is not available in its buffer.  However, the missing transmission from node 4 is available in its own buffer. Node 2 sends a G-NACK to its child nodes indicating that the first transmission from node 6 is missing. 
During the retransmission phase, nodes also overhear G-NACK messages to maintain local knowledge.  Although node 4 has the missing transmission in its buffer, it would abstain from accessing the first slot of the signaling channel as it is not free as per its local knowledge. The G-NACK  from the controller informs node 4 that node 2 (a higher priority node) has missing transmission(s). Further, the G-NACK  overheard from node 2 informs node 4 that the missing transmission from node 2 is not in its own buffer. Hence, node 4 ascertains that the first two RFS slots can potentially be occupied as node 2 would request timeslots for itself in case the missing transmission is present in the buffer of its child node. 
Similar to the previous case, we assume that the missing transmission from node 6 is available in its own buffer. As per Rule RS-1, it requests timeslots \(t_0\) and \(t_1\) in an RFS-RS message  in slot \(s_0\). Node 2 confirms the requested allocation in slot \(s_1\). As per Rule RS-2, node 2 requests timeslot \(t_2\) by sending an RFS-RS message in slot \(s_2\). The controller confirms the requested allocation in slot \(s_3\). In slot \(s_4\), node 4 requests timeslots \(t_3\) and \(t_4\), which are the earliest available timeslots as per its local knowledge. The controller confirms the requested allocation and terminates the scheduling phase subsequently. Once the signaling phase is over, nodes switch to the retransmission channel. Note that  node 4 is the default relay for node 2 and vice versa, and node 7 is the default relay for node 6. First, the missing transmission of node 6 is recovered cooperatively. Timeslot \(t_0\) is used for parent-relay communication. In timeslot \(t_1\), node 7 and node 6 simultaneously transmit the missing information which leads to CI. Timeslot \(t_2\) is used by node 2 to forward node 6's information to the controller. Finally, timeslots \(t_3\) and \(t_4\) are used to recover node 4's missing information in a similar way.

\subsubsection{Case D}
This case is similar to Case B except that node 6 does not have the missing transmission in its buffer. This could happen if the downlink transmission from node 2 to node 6 failed. Consequently, node 6 did not transmit in the uplink.  After receiving the G-NACK from node 2, node 6 determines the need for downlink transmission. In this case, it would request three timeslots. The first timeslot (\(t_0\)) is used for parent-relay communication where node 2 will inform node 7 of the need for cooperative downlink transmission. In the second timeslot (\(t_1\)), both node 2 and node 7 simultaneously transmit the downlink information. The third timeslot (\(t_3\)) is used by node 6 to transmit its uplink information to node 2.

The retransmission scheduling technique  provides very high reliability. It also provides  flexibility as the retransmission schedule is customized for recovering the missing transmissions only. However, it  encompasses a non-deterministic signaling component and the  overhead of building a schedule.

\subsection{Schedule Extrapolation}
\begin{figure}
\centering
\includegraphics[scale=0.27]{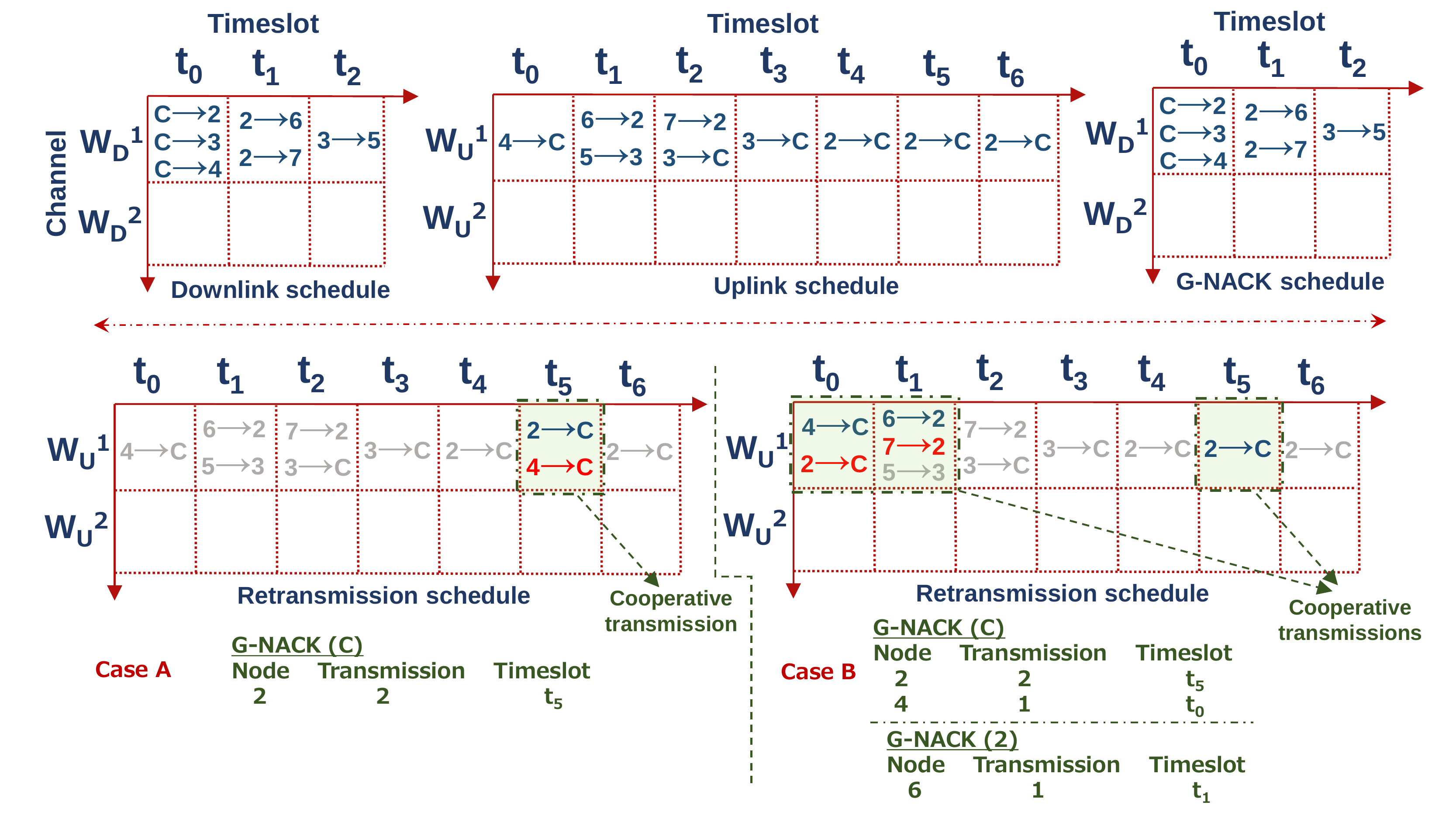}
\caption{\textcolor{black}{Schedule extrapolation technique and the retransmission schedule in different cases for the topology in Fig. \ref{ret_sch}. } }
\vspace{-0.5cm}
\label{sch_ext}
\end{figure}

The third technique integrates the key benefits of the first two techniques. It extrapolates the transmission schedule for recovering  missing transmissions through cooperative transmissions \emph{without building a retransmission schedule}. This is achieved by embedding some extra information in the G-NACKs which are disseminated through the downlink schedule. The need for retransmissions is ascertained by the controller. However,  retransmissions are handled based on the transmission schedule. 
The schedule extrapolation technique is  explained through the following cases which are illustrated in Fig. \ref{sch_ext}.

\subsubsection{Case A}
Consider the scenario wherein node 2's second transmission was missing at the controller at the end of the transmission schedule; however, it was available in node 2's buffer. The controller generates a G-NACK message indicating the missing transmission. Unlike retransmission scheduling technique, the controller also embeds the timeslot information of the missing transmission in the G-NACK message, which in this case is \(t_5\). The G-NACK from the controller also informs the relay node (node 4) to engage in a cooperative transmission at timeslot \(t_5\). Therefore, in timeslot \(t_5\), both node 4 and node 2  simultaneously transmit the missing information to the controller which leads to CI. Note that a schedule for retransmission is not explicitly constructed. The original uplink schedule is duplicated for recovering failed transmissions while exploiting cooperative transmissions for enhanced reliability. It is important to mention that the successful transmissions in the previous round may not be retransmitted (duplicated). 

\subsubsection{Case B}
Consider that node 2's second transmission and the only transmission of node 4 were missing at the controller at the end of the transmission schedule. The controller generates a G-NACK message indicating these transmissions and the respective timeslots. We assume that the missing transmission from node 4 was available while that of node 2 was not available in their respective buffers. Hence, node 2 generates a G-NACK message indicating the missing transmission of node 6 and the associated timeslot. The G-NACK from the controller informs nodes 2 and 4 to engage in cooperative transmissions for each other in timeslots \(t_0\) and \(t_5\), respectively. The G-NACK from node 2 informs node 7 to engage in a cooperative transmission for node 6 in timeslot \(t_1\). Note that any relay node would not engage in a cooperative transmission if it does not have the missing information in its buffer. In timeslot \(t_0\) of the retransmission schedule, node 4 and node 2 simultaneously transmit the missing information which leads to CI. Similarly, node 6 and node 7 engage in a cooperative transmission in timeslot \(t_1\). Finally, node 2 forwards node 6's information to the controller in timeslot \(t_5\). Note that, node 2 can also forward the missing information based on the aforementioned SNC operation.

The schedule extrapolation technique achieves very high reliability while providing deterministic performance. However, it may not provide the same level of flexibility as the retransmission scheduling technique. 

\begin{figure}
\centering
\includegraphics[scale=0.31]{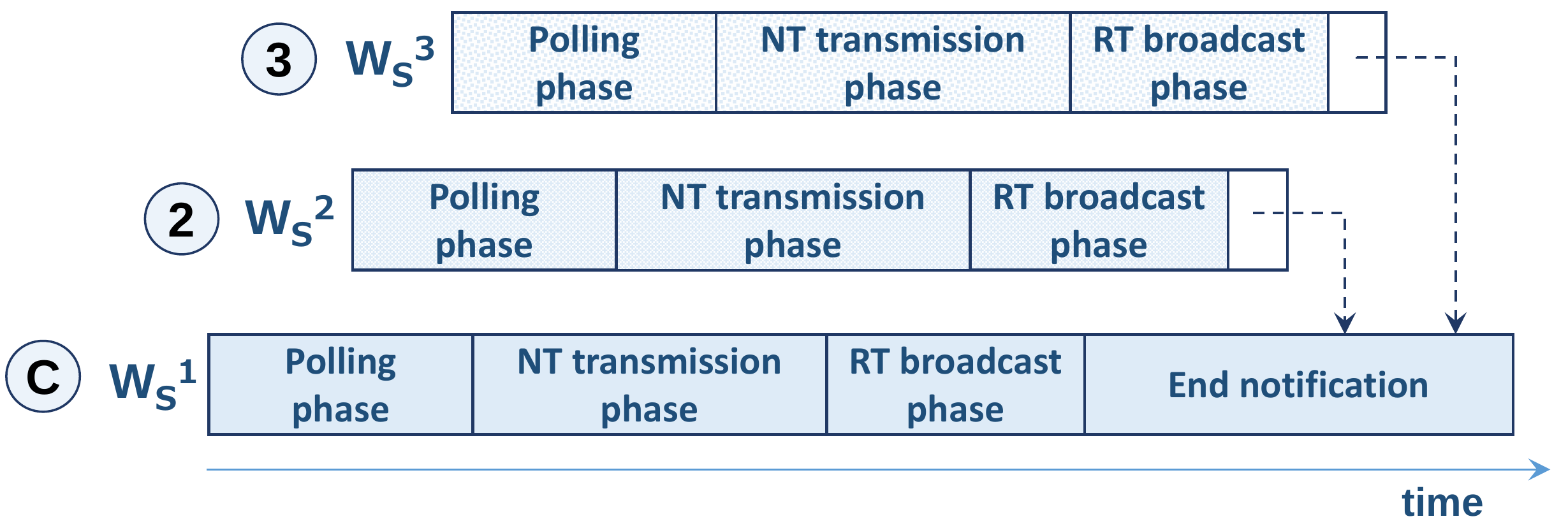}
\caption{\textcolor{black}{Relay selection technique in \textsf{GALLOP}  (same topology as Fig. \ref{dist_sch1}).}  }
\vspace{-0.5cm}
\label{rel_sel}
\end{figure}

\subsection{Relay Selection}
\textsf{GALLOP} implements a simple relay selection technique (illustrated in Fig. \ref{rel_sel}) which is based on link-level measurements. It begins by the controller polling each of its child nodes by transmitting an empty RFS message on the default signaling channel. Each child node responds with an empty ASGN message. The neighboring child nodes of the controller which successfully receive the ASGN message add the ID of the polled child along with the received signal-to-noise ratio (SNR) in a neighbor state table (NT). Each child node only records \(N_b\) strongest neighbors. After successfully polling all child nodes, the controller transmits a DLS message which contains a schedule for its child nodes on the default signaling channel. Based on the schedule information, each child node reports its NT to the controller. Upon receiving the NTs, the controller computes a relay state table (RT) that includes a relay node for each child node. The controller picks the relay node based on the highest SNR from every child's NT. It broadcasts the RT on the default signaling channel. 
Each
non-leaf child node repeats the aforementioned operation for its 
child nodes on a different channel after receiving the RT from the controller. For parallel operation of multiple branches, the controller allocates a subset
of orthogonal channels to all of its child nodes. The channel information can potentially be
embedded in the schedule information for receiving the NTs. The
procedure continues in the downward direction such that each parent
node allocates a subset of orthogonal channels to its child nodes.
A leaf node notifies its parent by transmitting
an uplink signaling message that indicates end of relay selection operation,
on the respective channel. The relay selection process completes once the
controller has received an uplink signaling message from all of its
child nodes.
Relay selection procedure can also be realized
in an online manner based on overhearing during data transmission
phases. If a node successfully overhears from a neighbor (with the same default parent), with an average SNR (over a fixed number of phase)  above a certain threshold, it notifies its parent to act as a relay for the neighbor. The information exchange pertaining to relay selection can be piggybacked on uplink/downlink transmissions.



\section{\textsf{GALLOP} -- Convergence Time Analysis} \label{sect_ana}
We adopt an SNR-based model for link outage, where a transmission is successful if the instantaneous SNR exceeds a threshold. We consider the case of an arbitrary parent node \(p\) with \(K\) directly connected child nodes uniformly distributed in its coverage radius, denoted by \(R_p\). Let \(\gamma_{\text{sig}}=\mathcal{P}^t \vert h \vert^2 d^{-\alpha}/ \sigma^2\) denote the instantaneous SNR of a signaling transmission in either downlink or uplink such that \(\mathcal{P}^t\) denotes the transmit power, \(h\) denotes the channel fading coefficient, \(d\) denotes the distance, \(\alpha\) is the path loss exponent, and \(\sigma^2\) is the noise power. Let, \(P_o^{\text{sig}}\) denote the average outage probability for the signaling transmission which can be evaluated as 
\begin{equation}
P_o^{\text{sig}}=\mathbb{E}_d \left[ \mathbb{P} \left\lbrace\gamma\leq \beta \right\rbrace \right]
 =\mathbb{E}_d \left[ \mathbb{P}  \left\lbrace\vert h \vert^2 \leq \beta \sigma^2 d^{\alpha} / \mathcal{P}^t \right\rbrace \right],
\end{equation}
where \(\beta\) is the threshold SNR for successful transmission and \(\mathbb{E}_d\) denotes the expectation over the distance. The probability density function (PDF) of the distance between the parent node and the \(k\)th child node is given by \cite{dist_reg}
\begin{equation}
\label{pdf_dist_1}
f(d,k)=\frac{\left(1-d^2/R_{p}^2\right)^{K-k} \left( d^2/R_{p}^2\right)^{k-1}}{B\left( K-k+1, k\right)} \frac{2d}{R_{p}^2},
\end{equation}
where \(B(.)\) is the Beta function. Using \eqref{pdf_dist_1} and under the assumption of widely used Rayleigh fading, \(P_o^{\text{sig}}\) is given by
\begin{equation}
\label{o_dl_a}
\begin{aligned}
\allowdisplaybreaks
P_o^{\text{sig}}&=\frac{2}{R_{p}^2}\int_0^{R_{p}} d \left[ 1-\exp \left(-\zeta  d^\alpha \right)  \right] \mathrm{d}d, 
\end{aligned}
\end{equation}
where \(\zeta= \beta \sigma^2/\mathcal{P}^t\). 
We define \(P_a^f\) as the probability of allocation failure for a node which is given by 
\begin{equation}
\label{fail_alloc}
P_a^f=\left[ P_o^{\text{RFS}}+\left(1-P_o^{\text{RFS}} \right) P_o^{\text{ASGN}} \right] \cdot \left(1-P_o^{\text{DLS}}\right),
\end{equation}
where \(P_o^{\text{RFS}}\), \(P_o^{\text{ASGN}}\) and \(P_o^{\text{DLS}}\) denote the average outage probability for RFS, ASGN and DLS messages, respectively, which can be calculated using \eqref{o_dl_a}.


Next, we find the average convergence time. In the ideal scenario of no link-level transmission failures, the convergence time (in terms of signaling slots) for constructing a schedule is given by \(\mathcal{T}_{\text{conv}}^{I}=3K+1\). Let, \(w_o^k\) denote the access delay for the first RFS retransmission of the \(k\)th device. Then, the average access delay (based on signaling message retransmission policy) for successful allocation at first RFS retransmission can be calculated as
\begin{equation}
\begin{aligned}
\allowdisplaybreaks
\label{m_delay_1}
\mathbb{E} \left(w_o^k  \right)&=\frac{3}{K} P_a^f \left(1-P_a^f\right) \sum_{i=1}^K  \left[\left(i-1 \right) - i  \left(1-P_a^f \right) \right] \\
&=\frac{3}{2}P_a^f \left(1-P_a^f\right) \left[\left(K-1 \right) - \left(K+1 \right) \left(1-P_a^f \right)  \right].
\end{aligned}
\end{equation}

The average access delay for successful allocation at second or higher RFS retransmission can be calculated as
\begin{equation}
\begin{aligned}
\allowdisplaybreaks
\label{m_delay_2}
\mathbb{E} \left(w_r^k  \right)&=3\sum_{j=2}^{\infty} j  \left(P_a^f\right)^j \left(1-P_a^f\right) \left[  \frac{1+\Psi}{2} + \frac{1}{K}\sum_{i=1}^K  \left(i-1 \right) \right]\\
&=\frac{3\left(K+\Psi\right)}{2} \left[	\frac{2\left(P_a^f \right)^2-\left(P_a^f \right)^3}{1-P_a^f} \right].
\end{aligned}
\end{equation}

Hence, the mean convergence time for an arbitrary parent node \(p\) with \(K\) directly connected child nodes is given by 
\begin{equation}
\label{m_ct_p}
\mathcal{T}_{\text{conv}}=\mathcal{T}_{\text{conv}}^{I}+\mathbb{E} \left(w_o^k  \right)+\mathbb{E} \left(w_r^k  \right),
\end{equation}
where \(\mathbb{E} \left(w_o^k  \right)\) and \(\mathbb{E} \left(w_r^k  \right)\) are given by \eqref{m_delay_1} and \eqref{m_delay_2}, respectively. Note that \(\mathcal{T}_{\text{conv}}\) reduces to \(\mathcal{T}_{\text{conv}}^{I}\) when \(P_a^f =0\), which is intuitive. 
The proposed framework can be integrated with spatial modeling techniques to achieve bounds on the overall convergence time of a multi-hop topology. As this is a complete topic in itself, it has been left as part of  future work.


\begin{figure}
\centering
\includegraphics[scale=0.39]{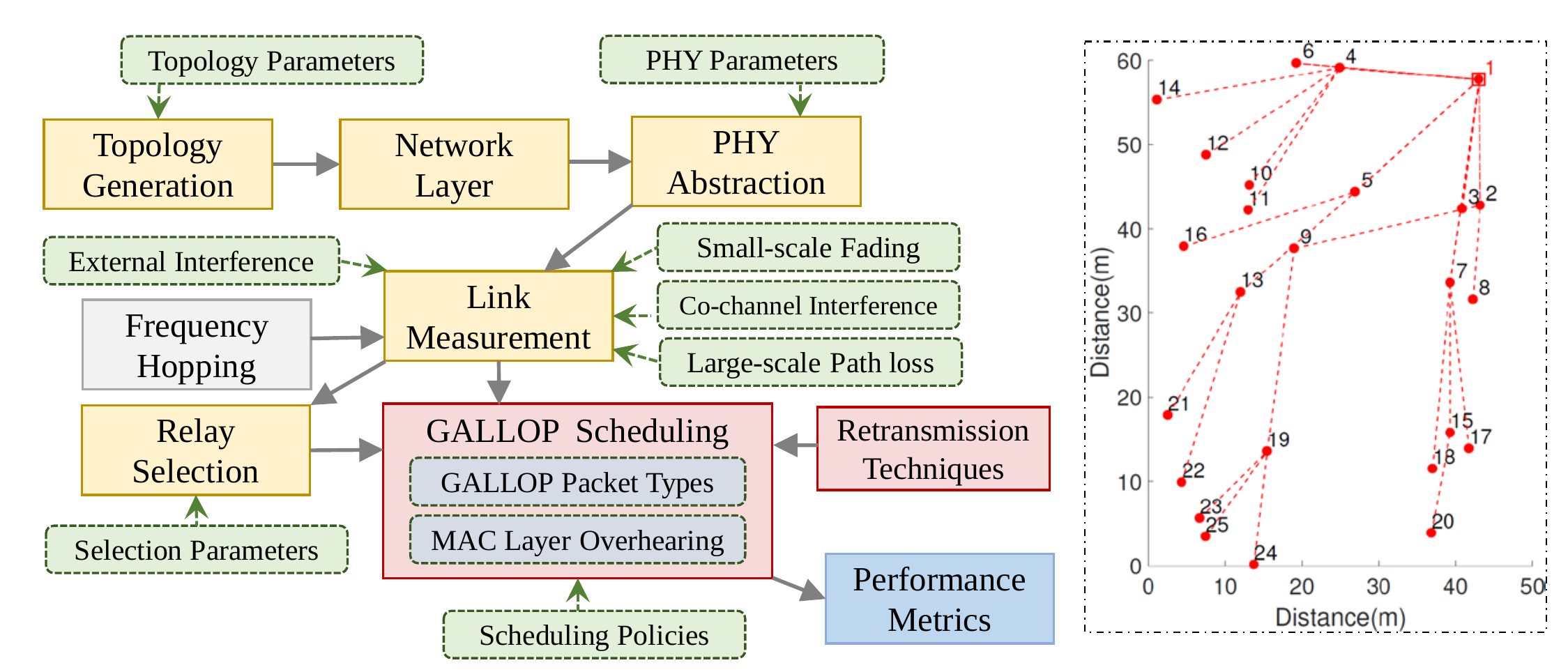}
\caption{Schematic block diagram of the \textsf{GALLOP} simulator.  }
\vspace{-0.3cm}
\label{sim}
\end{figure}

\begin{center}
	\begin{table}
		\caption{Parameters for Performance Evaluation}
		\begin{center}
			\begin{tabular}{ll|ll}
				\hline	
				\toprule
				Parameter  & Value & Parameter  & Value \\\hline
				\midrule
				Noise Power   & \(-174 \ \text{dBm/Hz}\) & Channel bandwidth & \(2 \ \)MHz \\
				Transmit power   & \(9 \ \text{dBm}\) & Path loss exponent (\(\alpha\))   & \(3.3\) \\
								Slot duration   & \(200\) \(\mu\)s  & PHY layer   & LE 2M \\
				\hline
			\end{tabular}
		\end{center}
		\vspace{-0.3cm}
			\label{params}
	\end{table}
\end{center}

\begin{figure}
	\centering
	\subfloat[]{\includegraphics[scale=0.2]{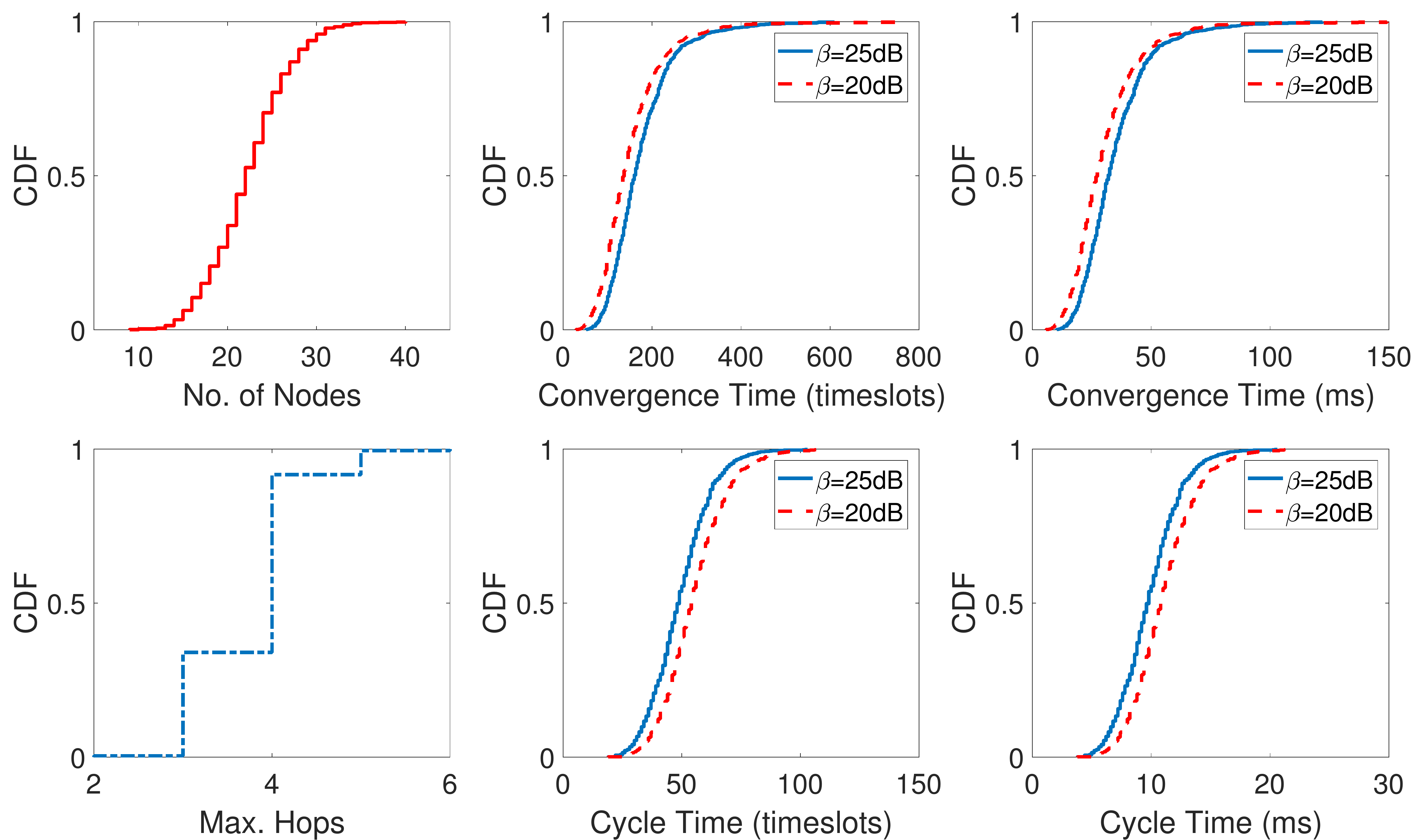}}\\
	\subfloat[]{\includegraphics[scale=0.2]{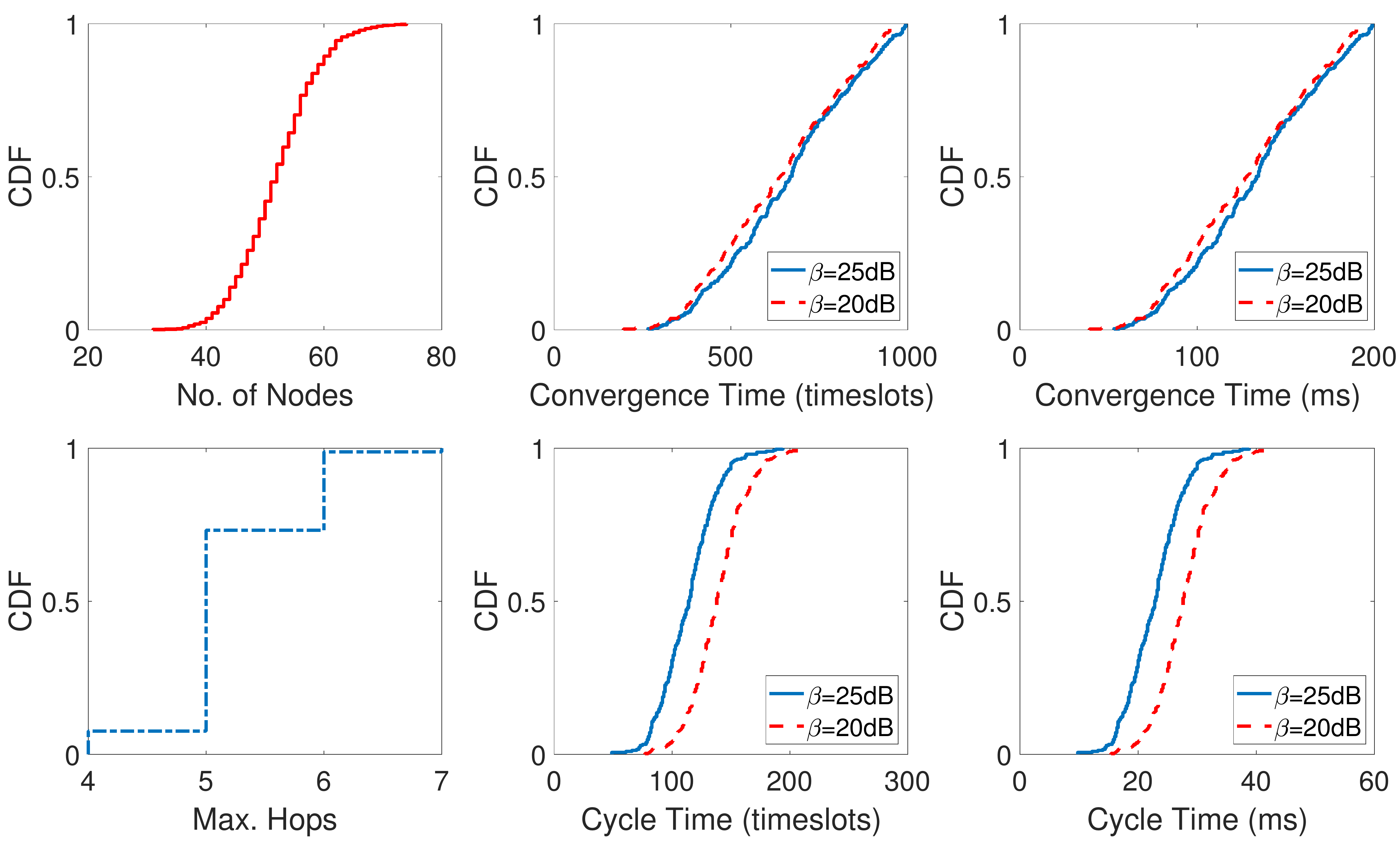}}\\
	\subfloat[]{\includegraphics[scale=0.21]{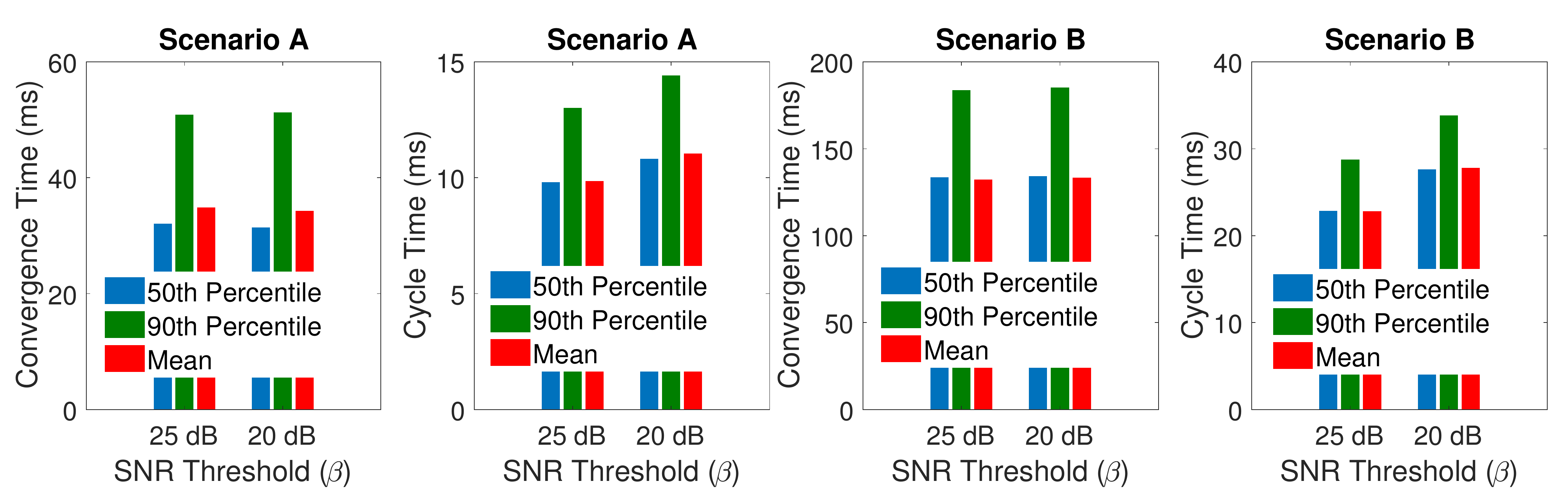}}\\
	\caption{Convergence time and cycle time performance for (a) \emph{Scenario A} and (b) \emph{Scenario B}. The  statistics are shown in (c).}
	\vspace{-0.5cm}
	\label{CC1}
\end{figure}

\section{\textsf{GALLOP} -- System-Level Simulations} \label{sect_perf}
\textcolor{black}{We have developed a MATLAB-based simulator for system-level evaluation of \textsf{GALLOP}. The schematic  of the simulator is shown in Fig. \ref{sim}.} We deploy a Poisson\footnote{\textcolor{black}{We have used Poisson distributed topology as the Poisson point process is the most widely used spatial point process for modeling connectivity in wireless (cellular, ad-hoc, mesh, etc.). networks. }} distributed mesh topology of nodes in a region, which is transformed into a tree
topology based on a routing protocol. 
We have implemented a simplified version of RPL \cite{RPL} at the network layer such that each node is assigned a rank based on the Euclidean distance from the controller. A realization of the multi-hop topology is  shown in Fig. \ref{sim}. The PHY abstraction block creates a PHY layer model based on various parameters. The link measurement block captures the channel model and accounts for small-scale Rayleigh fading, large-scale path loss, co-channel interference (collisions), and external interference. We adopt a well-known  path loss model, commonly employed for Bluetooth, wherein
the path loss (in dB) at a distance \(d\) is given by \(PL(d) =
PL(d_0) + 10 \alpha \log(d/d_0)\), such that \(PL(d_0) = 58.1\) dB,
\(d_0\) = \(8\) meters (m) and \(\alpha\) is the path loss exponent. The relay selection block implements the proposed relay selection technique. We have implemented \textsf{GALLOP} at the message level.  The slot duration  is set to \(200\) \(\mu\)s as in our hardware implementation. Other parameters are given in \tablename~\ref{params}.
We consider two scenarios for network topology and conduct Monte Carlo simulations
on different multi-hop topologies.  In \emph{Scenario A}, nodes are Poisson distributed with a mean value of \(20\) in a square region of side  \(60\) m. In \emph{Scenario B}, nodes are Poisson distributed with a mean value of \(50\) in a square region of side  \(80\) m.

Fig. \ref{CC1} shows the convergence time (which is defined as the time taken to build the schedule) and the cycle time for both scenarios based on \(1000\) iterations, with a different topology in each iteration. In \emph{Scenario A}, where \(10\) to \(40\) nodes are  distributed with a maximum of \(2\) to \(6\) hops from the controller, \textsf{GALLOP} achieves a mean convergence time of \(36\) ms (\(180\) slots) with a \(90\)th percentile of \(50\) ms (\(250\) slots), for an SNR threshold (\(\beta\)) of \(25\) dB.  In \emph{Scenario B}, where \(30\) to \(75\) nodes are distributed with a maximum  of \(4\) to \(7\) hops from the controller, the mean convergence time and the \(90\)th percentile  is \(130\) ms and \(182\) ms, respectively, for \(\beta=25\) dB. The convergence time decreases in both scenarios for \(\beta=20\) dB due to less stringent link-level requirements for the success of signaling messages.

In \emph{Scenario A}, \textsf{GALLOP} achieves a mean cycle time of \(9.5\) ms (\(90\)th percentile of \(13\) ms), for \(\beta=25\) dB. With a similar \(\beta\), the mean cycle time is \(22\) ms (\(90\)th percentile of \(29\) ms) in \emph{Scenario B}. The cycle time increases in both scenarios when \(\beta=20\) dB. This is because a lower value of \(\beta\) reflects a relatively denser environment. Hence, there are less parallel transmission opportunities which leads to increase in length of the overall schedule. Results demonstrate that \textsf{GALLOP} adapts the schedule as per network conditions due to local overhearing. In dense networks, it builds a sequential schedule for avoiding conflicts. In sparse networks, it exploits opportunities for parallel transmissions.

\begin{figure}
\centering
\includegraphics[scale=0.2]{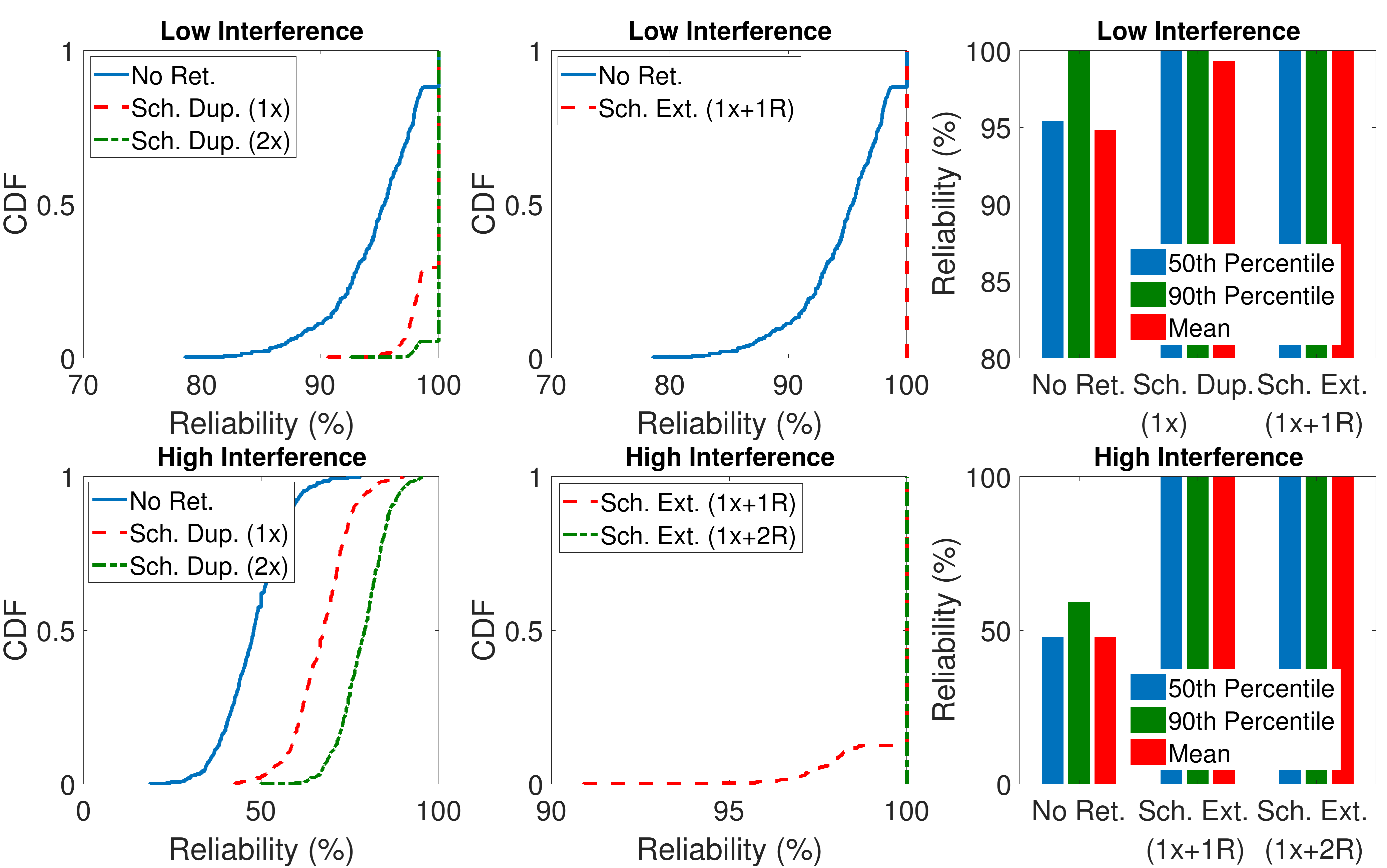}
\caption{Reliability performance of different techniques for \emph{Scenario A}. }
\vspace{-0.4cm}
\label{Rel1}
\end{figure}

Next, we evaluate the reliability (in terms of packet delivery) performance   and investigate the robustness of different retransmission techniques. We implement a spatio-temporal model for external Wi-Fi interference. From a spatial perspective, the distance of a receiver from an interfering Wi-Fi access point (AP) is uniformly distributed between \(1\) m and \(25\)  m. From a temporal perspective, the traffic of an AP is modeled as a two-state independent and identically distributed random process wherein the duration
of busy (ON) and idle (OFF) periods is  exponentially
distributed with a mean of \(\mu_{\text{ON}}\) and \(\mu_{\text{OFF}}\), respectively. We consider two interference scenarios. In \emph{low interference}, each link is interfered by a single AP, transmitting at \(14\) dBm with \(\mu_{\text{ON}}=0.25\) ms and  \(\mu_{\text{OFF}}=0.5\) ms. In \emph{high interference}, each link is interfered by up to three APs, transmitting at \(20\) dBm with \(\mu_{\text{ON}}=1.5\) ms and  \(\mu_{\text{OFF}}=0.5\) ms.
%
Fig. \ref{Rel1} shows the reliability performance of different retransmission techniques for \emph{Scenario A}. Under low interference, \textsf{GALLOP} achieves a mean reliability of \(94.79\%\) when no retransmissions are enabled. However, schedule duplication as well as schedule extrapolation  improve reliability. With one and two schedule duplications, the mean reliability increases to \(99.19\%\) and \(99.8\%\), respectively, and with schedule extrapolation to \(100\%\). In high interference, the mean reliability is \(48\%\) (\(90\)th percentile of \(59\%\)) when no retransmissions are enabled.  With one and two schedule duplications, the mean reliability increases to \(67\%\) (\(90\)th percentile of \(76.09\%\)) and \(78.7\%\) (\(90\)th percentile of \(86.55\%\)), respectively. We limit our evaluation to cooperative transmission 1. With schedule extrapolation (employing cooperative transmission 1), the mean reliability is \(99.68\%\) (\(90\)th percentile of \(100\%\)). The mean reliability with schedule extrapolation increases to \(100\%\) if up to two relay nodes are used. The results demonstrate the effectiveness of cooperative transmissions for very high reliability. Similar reliability gain has been achieved in \emph{Scenario B}. 
%
%
\begin{figure}
\centering
\includegraphics[scale=0.2]{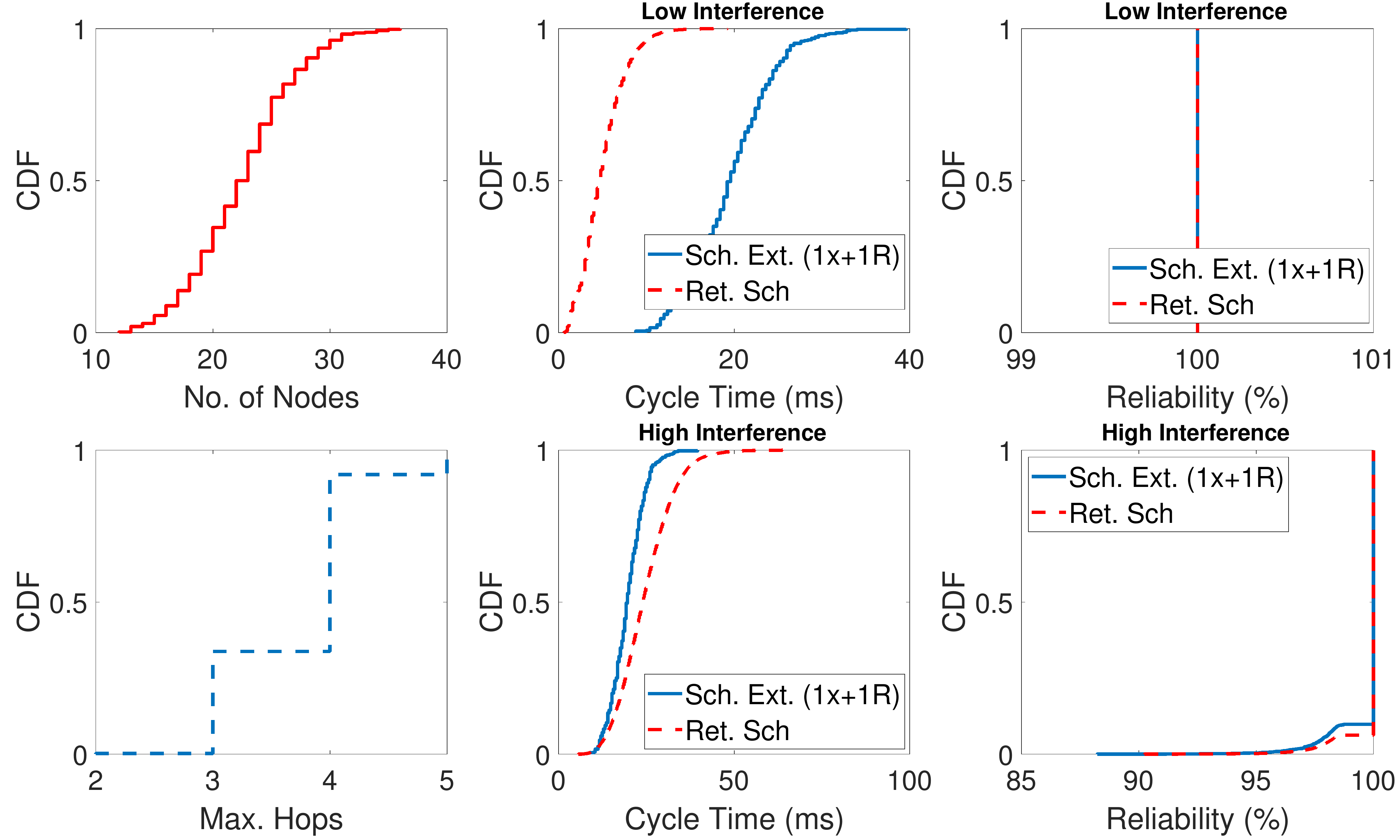}
\caption{Performance comparison of  retransmission scheduling and schedule extrapolation techniques for \emph{Scenario A}. }
\vspace{-0.4cm}
\label{Rel_rtsig}
\end{figure}
Fig. \ref{Rel_rtsig} compares the performance of retransmission scheduling and schedule extrapolation. Both techniques achieves very high reliability under low as well as high interference. The main difference comes in cycle time performance. Schedule extrapolation achieves a mean cycle time of \(19.8\) ms (\(99\) timeslots) with a \(90\)th percentile of \(26.8\) ms (\(134\) timeslots). In low interference, retransmission scheduling achieves a mean cycle time of \(5\) ms (\(25\) timeslots) with a \(90\)th percentile of \(8.4\) ms (\(42\) timeslots). However, under high interference, the mean cycle time increases to  
\(24.8\) ms (\(124\) timeslots) with a \(90\)th percentile of \(35\) ms (\(175\) timeslots). This increase is due to higher convergence time for constructing the retransmission schedule. Retransmission scheduling is  \(75\%\) better on average in terms of cycle time under low interference. However, under high interference, it performs \(20\%\) worse on average in terms of cycle time.

%

\begin{figure}
\centering
\includegraphics[scale=0.17]{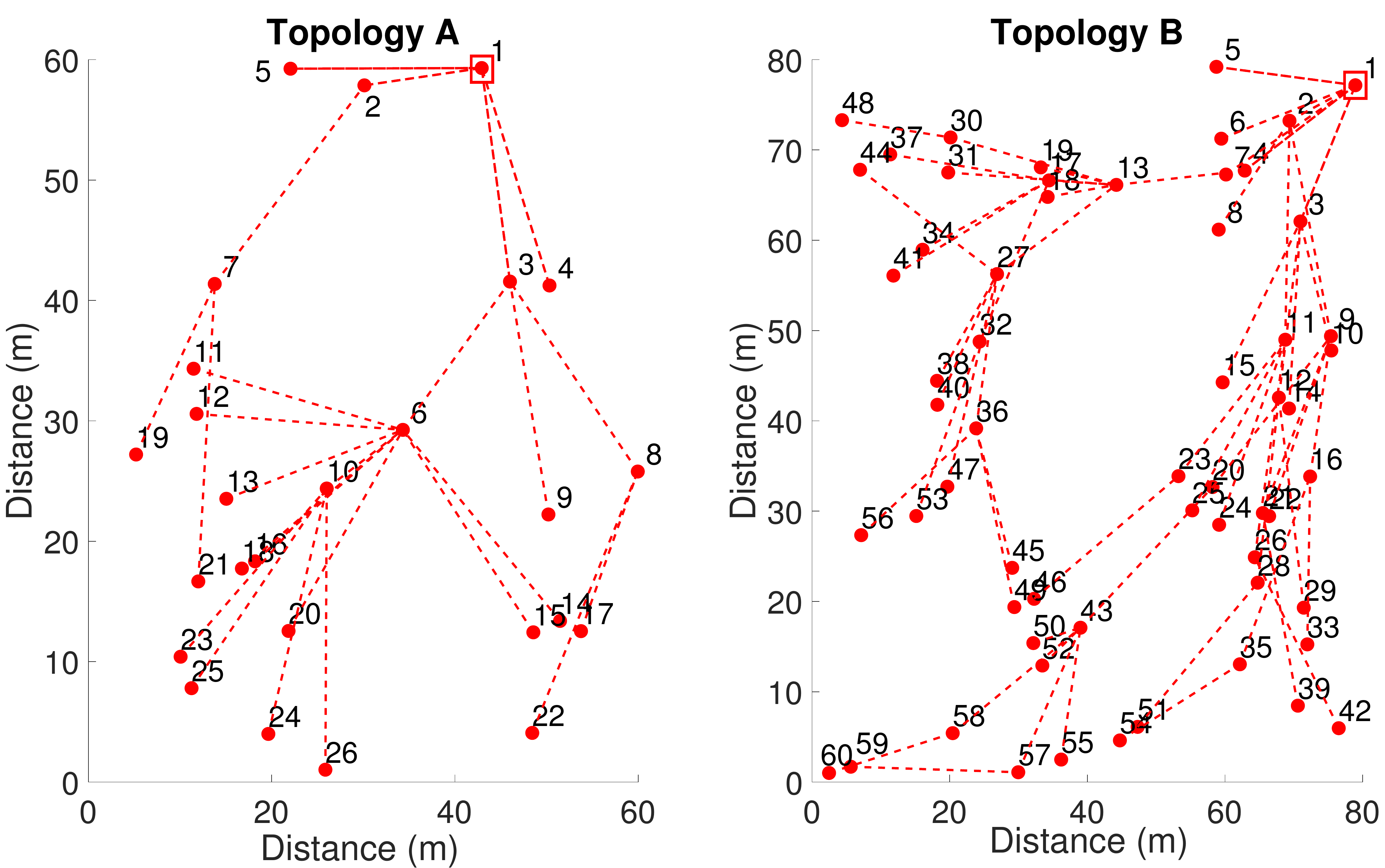}
\caption{Topology A and Topology B. Node 1 is the controller. }
\label{top_single}
\end{figure}

\begin{figure}
\centering
\includegraphics[scale=0.21]{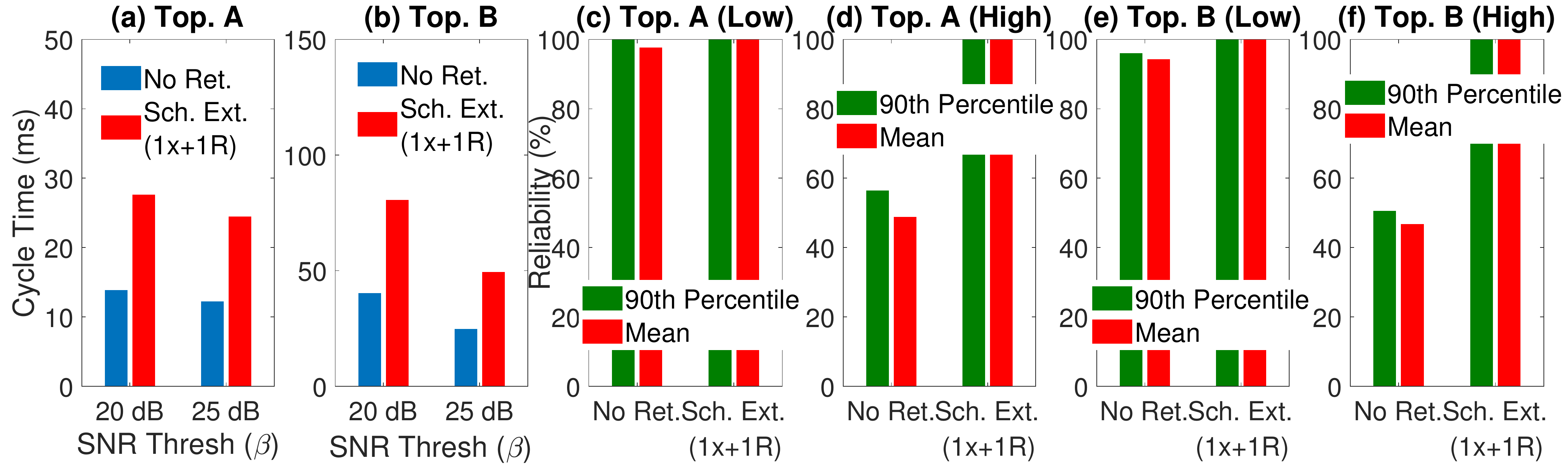}
\caption{Cycle time (in (a) -- (b)) and reliability (in (c) -- (f)) performance for Topologies A and B. }
\vspace{-0.5cm}
\label{top_single_perf}
\end{figure}

We also evaluate the performance of \textsf{GALLOP} on two distinct topologies  shown in Fig. \ref{top_single}. Topology A comprises \(26\) nodes with maximum of \(4\) hops whereas Topology B consists of \(60\) nodes with maximum of \(6\) hops. The cycle time and reliability performance for both topologies is shown in Fig. \ref{top_single_perf}. Key results are summarized as follows. 
 For \(\beta=25\) dB, the cycle time for Topology A is \(12.2\) ms with no retransmissions. The cycle time doubles to \(24.4\) ms with schedule extrapolation. In a similar setting, the cycle time for Topology B is  \(24.6\) ms with no retransmissions and and \(49.2\) ms with schedule extrapolation. 
In low interference, the mean reliability with no retransmissions is \(97.2\%\) (\(90\)th percentile of \(100\%\)) and \(94.15\%\) (\(90\)th percentile of \(95.97\%\)) for Topology A and Topology B, respectively. With schedule extrapolation, the mean reliability is \(100\%\). 
In high interference, the mean reliability with no retransmissions is \(48.6\%\) (\(90\)th percentile of \(56.3\%\)) and \(46.69\%\) (\(90\)th percentile of \(50.5\%\)) for Topology A and Topology B, respectively. Using schedule extrapolation, the mean reliability is \(99.97\%\) for Topology A and \(99.83\%\) for Topology B, respectively. 

%
%
%
%
%
%
%
%
%
%


\begin{figure}
\centering
\includegraphics[scale=0.21]{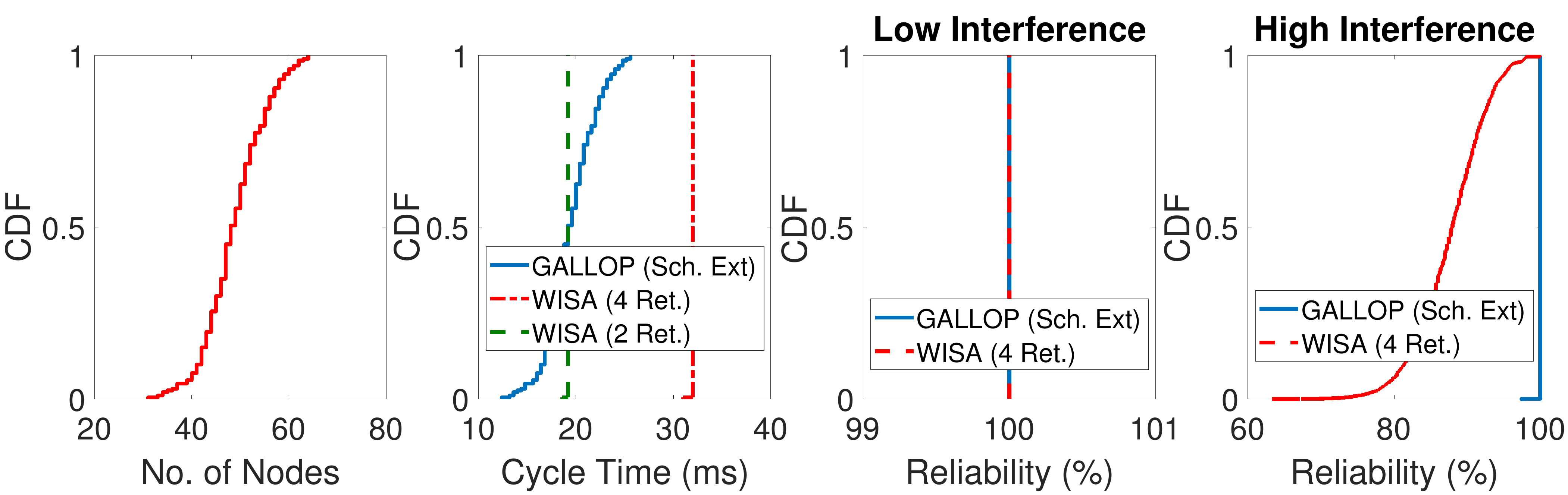}
\caption{Singe-hop performance of \textsf{GALLOP} (\(\beta=25\) dB). }
\vspace{-0.4cm}
\label{sh_perf}
\end{figure}

 \textsf{GALLOP} is equally capable of single-hop operation. Its distributed scheduling algorithm is agnostic to the nature of topology. Fig. \ref{sh_perf} shows the single-hop performance and its benchmarking against WISA. In a single-hop network comprising \(30\) -- \(65\) nodes within controller's coverage, the mean cycle time of \textsf{GALLOP} with schedule extrapolation  is \(98\) timeslots or \(19.6\) ms (\(90\)th percentile of \(114\) timeslots or \(22.8\) ms). On the other hand, WISA provides a cycle time of \(160\) timeslots (\(32\) ms) and \(96\) timeslots (\(19.2\) ms) with \(4\) and \(2\) mandatory retransmissions, respectively. Both technologies achieve similar reliability of \(100\%\) under low interference. Under high interference, WISA with \(4\) retransmissions achieves a mean reliability of \(87.6\%\) (\(90\)th percentile of \(93.75\%\)), whereas \textsf{GALLOP} achieves a mean reliability of \(99.94\%\) (\(90\)th percentile of \(100\%\)). Results demonstrate that \textsf{GALLOP} outperforms WISA, especially in terms of reliability performance while providing similar or better latency performance. 
%
%
%

\begin{figure}
\centering
\includegraphics[scale=0.2]{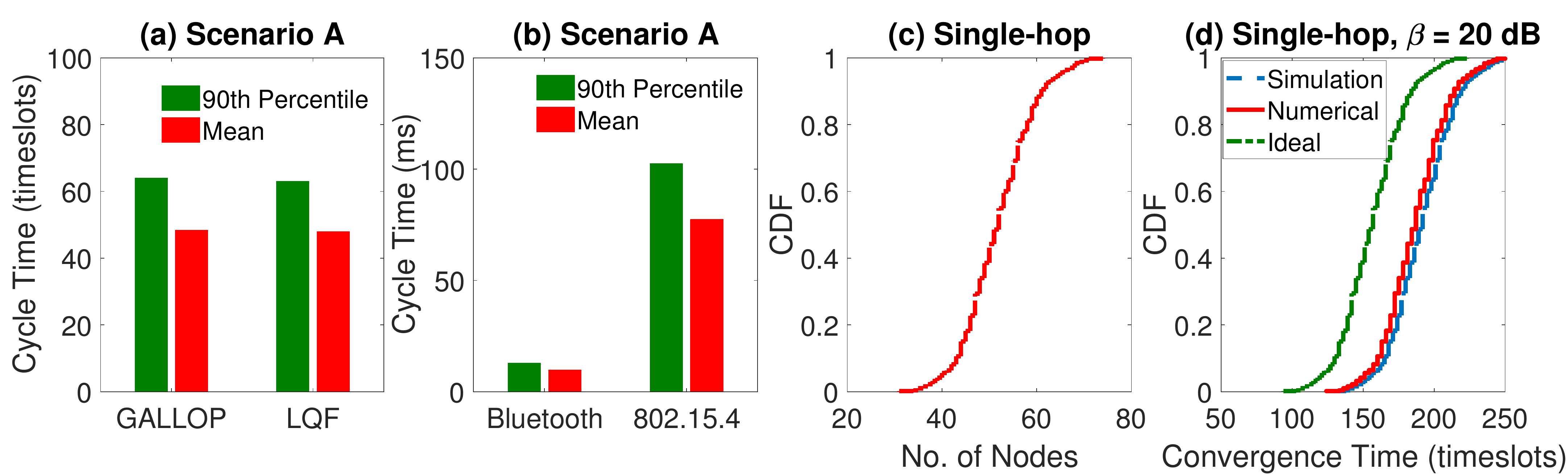}
\caption{\textsf{GALLOP} evaluation from (a) algorithmic and (b) hardware aspects. Scenario and results for convergence time  are shown in (c) and (d).  }
\vspace{-0.5cm}
\label{ta_comp}
\end{figure}

Fig. \ref{ta_comp} captures the performance of \textsf{GALLOP} from different aspects. First, it evaluates the efficiency of distributed scheduling which is  benchmarked  against the \emph{longest queue first} (LQF) algorithm \cite{LQF_algo}, also known as \emph{greedy maximal scheduling}.  LQF is a centralized algorithm and often used as a heuristic for  optimal performance. As per LQF, nodes are sorted in a descending order with respect to their queue lengths. The node with the longest queue is allocated the current timeslot. Other nodes  which can transmit without causing conflict are also allocated the same timeslot. For \emph{Scenario A}, LQF provides a \emph{schedule duration} of \(47.9\) timeslots with a \(90\)th percentile of \(62.8\) timeslots whereas \textsf{GALLOP} provides a mean cycle time of \(48.4\) timeslots with a \(90\)th percentile of \(64.5\) timeslots. The results demonstrate that distributed scheduling  of \textsf{GALLOP} provides a near-optimal performance despite its practicality and simplicity. Note that LQF is centralized in nature and requires  queue length information for all the nodes in the network.  LQF is also not control-aware like \textsf{GALLOP}. While LQF calculates an optimal schedule for a given topology, it cannot ensure minimal cycle time. 
Fig. \ref{ta_comp} also shows that \textsf{GALLOP} would provide at least \(8\) times better performance, in terms of minimum cycle time, on a LE 2M PHY  as compared to IEEE 802.15.4 PHY which is used by WirelessHART. Fig. \ref{ta_comp}  also shows the convergence time performance based on numerical and simulation results. The simulation results closely follow the numerical results and validate the accuracy of modeling convergence time.

\section{\textsf{GALLOP} -- Hardware Implementation} \label{sect_imp}
We have implemented \textsf{GALLOP} on the Nordic Semiconductor nRF52840-PDK platform (\url{https://www.nordicsemi.com/eng/Products/nRF52840}) which is built around a 32-bit ARM Cortex-M4F CPU with 1 MB flash and 256 kB RAM on chip. The embedded 2.4 GHz transceiver supports multiple protocols. \textsf{GALLOP} has been implemented on the LE 2M PHY of Bluetooth 5. However, it has been tested on the LE Coded PHY (125 kbps) as well. 
Our \textsf{GALLOP} implementation adopts an enhanced version of Glossy \cite{rob_flood} and the Atomic stack \cite{atomic} ported to the nRF52840 radio for multi-hop time synchronization. Following time synchronization, \textsf{GALLOP} initiates a time-slotted operation that is strictly enforced. The state machine is driven by radio and timer interrupts. It can change on three distinct events: packet successfully received, timeslot start and packet transmitted. The transmission decisions are handled on-the-fly at the end of every slot based on current state and packets queued for  next available slot of the correct type. Each transmit slot is partitioned into pre-processing and transmission preparation, radio ramp-up and transmission and post-processing and protocol state operation. For each device, \textsf{GALLOP} uses an 8-bit ID which is generated by a hash function of its address.  

\begin{figure}
\centering
\includegraphics[scale=0.45]{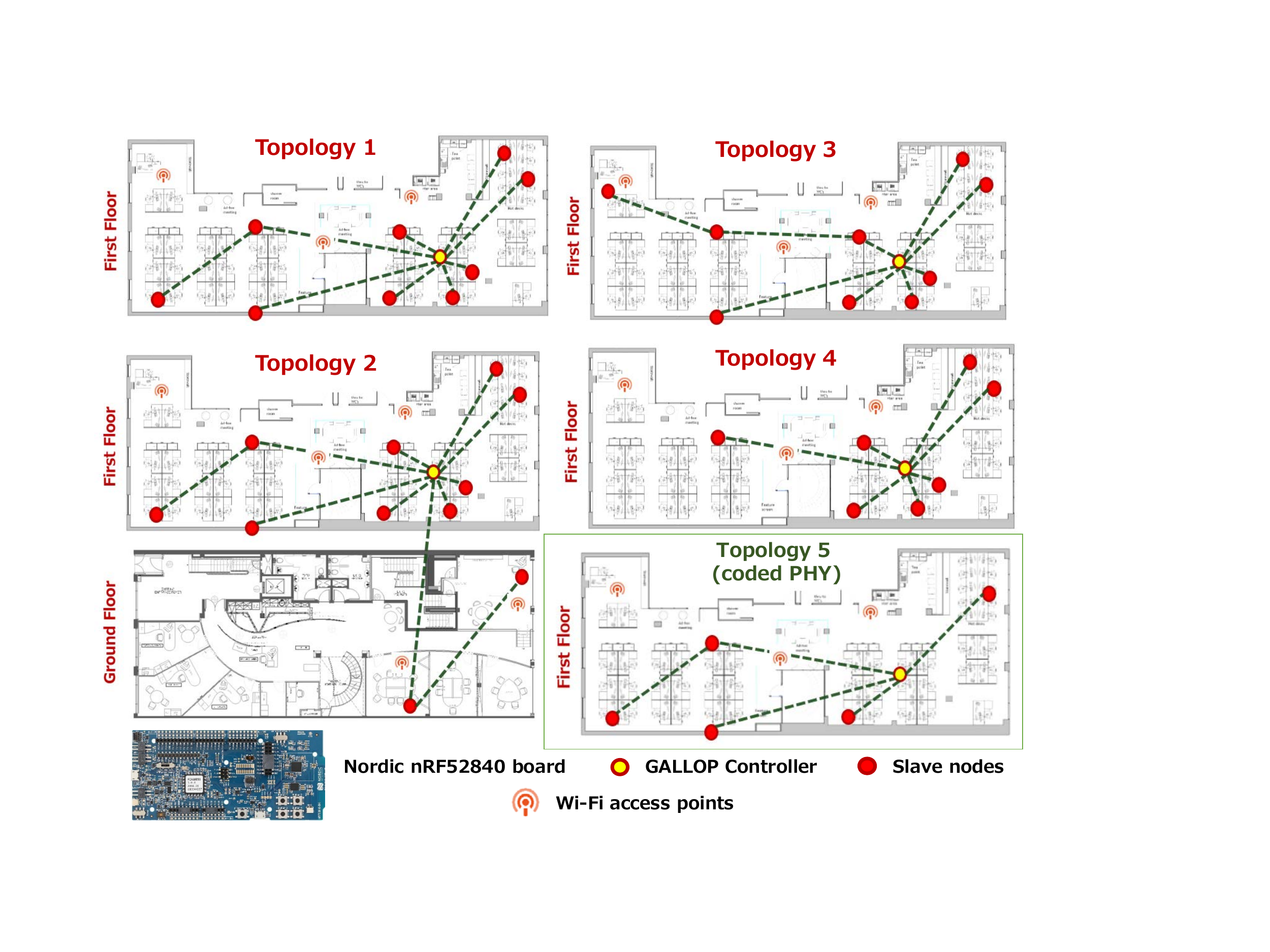}
\caption{Testbed topologies for multi-hop and single-hop performance evaluation of \textsf{GALLOP}.   }
\vspace{-0.4cm}
\label{tops}
\end{figure}

We have conducted performance evaluation \textsf{GALLOP} over a testbed of Nordic nRF52840 devices deployed over two floors of our office (covering approximately \(600\) square meters). The testbed devices experience interference from Wi-Fi APs and other Bluetooth devices operating in office premises. We consider different multi-hop topologies  which are shown in Fig. \ref{tops}. Each topology consists of a single controller and multiple slave nodes. Topology 1 consists of \(9\) slaves with one \(2\)-hop connection. Topology 2 comprises  \(11\) slaves with two \(2\)-hop connections. Topology 3 consists of  \(9\) slaves with one \(3\)-hop connection and one \(2\)-hop connection. Topology 4 comprises  \(7\) slaves in single-hop connectivity. In addition to these topologies, there is a Topology 5 which consists of \(5\) slave nodes with one \(2\)-hop connection. Topology 5 is only used for testing the performance of LE Coded PHY. 

\subsection{Evaluation on LE 2M PHY}
We consider a bi-directional traffic model with cyclic information exchange where the controller sends command messages in the downlink and slave nodes respond back in the uplink. The \textsf{GALLOP} MAC protocol data unit (PDU) is \(16\) bytes in length. The overall packet size  is \(24\) bytes as the LE 2M PHY  has an  \(8\) byte overhead. The minimum required slot duration is \(136\) \(\mu\)s which accounts for a \(40\) \(\mu\)s transmit/receive turnaround time for the nRF52840 radio. We have fixed the timeslot duration to \(200\) \(\mu\)s to account for additional software delays. The transmit power is  \(9\) dBm.

We have evaluated the convergence time performance of  \textsf{GALLOP} based on \(3000\) iterations over our testbed. Key results are summarized in \tablename~\ref{res_conv}.
%
%
%
The results demonstrate the viability of  distributed scheduling  for both multi-hop and  single-hop topologies. The results also demonstrate the capability of fast convergence (on the order of few ms) for building a schedule. It is emphasized that the schedule for cyclic  exchange needs to be
built once for a specific topology. This is usually the case for most
industrial applications, even for highway platooning. However, \textsf{GALLOP} is capable of rapid reconfiguration in dynamic environments with frequent topological changes. 




\begin{table}[]
\caption{Summary of Testbed Evaluation (Convergence Time)}
\centering
\begin{tabular}{ccc}
\cline{1-3}
\toprule
\textbf{Topology} & \textbf{\begin{tabular}[c]{@{}c@{}}Convergence Time\\ (Mean)\end{tabular}} & \textbf{\begin{tabular}[c]{@{}c@{}}Convergence Time\\ (90th Percentile)\end{tabular}} \\
\midrule 
\midrule
Topology 1        & \begin{tabular}[c]{@{}c@{}}60.87 timeslots (12.17 ms)\end{tabular}       & \begin{tabular}[c]{@{}c@{}}63 timeslots (12.6 ms)\end{tabular}                      \\
Topology 2        & \begin{tabular}[c]{@{}c@{}}67.19 timeslots (13.43 ms)\end{tabular}       & \begin{tabular}[c]{@{}c@{}}69 timeslots (13.8 ms)\end{tabular}                      \\
Topology 3        & \begin{tabular}[c]{@{}c@{}}56.4 timeslots (11.28 ms)\end{tabular}        & \begin{tabular}[c]{@{}c@{}}60 timeslots (12 ms)\end{tabular}                        \\
Topology 4        & \begin{tabular}[c]{@{}c@{}}21.05 timeslots (4.21 ms)\end{tabular}        & \begin{tabular}[c]{@{}c@{}}21 timeslots (4.2 ms)\end{tabular}  \\  
\cline{1-3}                  
\end{tabular}
\vspace{-0.4cm}
\label{res_conv}
\end{table}


The cycle time and reliability performance of \textsf{GALLOP}, which is summarized in \tablename~\ref{t1}, has been evaluated over multiple phases of cyclic  exchange. \textsf{GALLOP} provides a cycle time of \(2.4\) ms for Topology 1,  \(2.6\) ms for Topology 2, \(3\) ms for Topology 3, and \(1.6\) ms for Topology 4. The reliability is measured in terms of mean packet delivery ratio (PDR) at the controller. In the absence of retransmissions and frequency hopping, the achieved reliability performance is not sufficient. For example, the achieved reliability for Topology 1 is \(95.27\%\). However, the reliability improves significantly with schedule extrapolation and frequency hopping. For instance, the reliability for Topology 1 increases to \(99.95\%\). Similar improvement in reliability has been achieved for other topologies as well.  The evaluation  reveals that cooperative transmissions are crucial for very high reliability. For example, in case of Topology 1, schedule extrapolation with no frequency hopping still achieves a reliability of \(99.73\%\). 
\begin{table}[]
\centering
\caption{Summary of Testbed Evaluation (LE 2M PHY)}
\begin{tabular}{ccccc}
\cline{1-5}
				\toprule
Topology                                                    &  Phases & Cycle Time                                                    & \begin{tabular}[c]{@{}c@{}}Successful \\ Packets\end{tabular} & PDR     \\ \cline{1-5} \midrule
\multicolumn{5}{c}{\textbf{No Retransmissions; No Frequency/Channel Hopping}}\\
\midrule
\begin{tabular}[c]{@{}c@{}}Topology 1\end{tabular} & 2013          & \begin{tabular}[c]{@{}c@{}}12 timeslots (2.4 ms)\end{tabular} & 17260 & 95.27\%
\\ 
\begin{tabular}[c]{@{}c@{}}Topology 2\end{tabular} & 1015          & \begin{tabular}[c]{@{}c@{}}13 timeslots (2.6 ms)\end{tabular} & 10551 & 94.5\%
\\
\begin{tabular}[c]{@{}c@{}}Topology 3\end{tabular} & 1100          & \begin{tabular}[c]{@{}c@{}}15 timeslots (3 ms) \end{tabular} & 9330 & 94.24\%
\\
\begin{tabular}[c]{@{}c@{}}Topology 4\end{tabular} & 1100          & \begin{tabular}[c]{@{}c@{}}8 timeslots (1.6 ms) \end{tabular} & 7538 & 97.9\%
\\ \cline{1-5} 
\midrule
\multicolumn{5}{c}{\textbf{Schedule Extrapolation with Frequency/Channel Hopping}}\\
\midrule 
\begin{tabular}[c]{@{}c@{}}Topology 1\end{tabular} & 2003          & \begin{tabular}[c]{@{}c@{}}24 timeslots (4.8 ms)\end{tabular} & 18018 & 99.95\%
\\ 
\begin{tabular}[c]{@{}c@{}}Topology 2\end{tabular} & 1000          & \begin{tabular}[c]{@{}c@{}}26 timeslots (5.2 ms)\end{tabular} & 10918 & 99.25\%
\\
\begin{tabular}[c]{@{}c@{}}Topology 3\end{tabular} & 1026         & \begin{tabular}[c]{@{}c@{}}30 timeslots (6 ms)\end{tabular} & 9190 & 99.52\%
\\
\begin{tabular}[c]{@{}c@{}}Topology 4\end{tabular} & 2006          & \begin{tabular}[c]{@{}c@{}}16 timeslots (3.2 ms)\end{tabular} & 13976 & 99.53\%
\\ \cline{1-5} 
\midrule
\multicolumn{5}{c}{\textbf{Schedule Extrapolation without Frequency/Channel Hopping}}\\
\midrule 
\begin{tabular}[c]{@{}c@{}}Topology 1\end{tabular} & 1000          & \begin{tabular}[c]{@{}c@{}}24 timeslots (4.8 ms)\end{tabular} & 8976 & 99.73\%
\\
\begin{tabular}[c]{@{}c@{}}Topology 4\end{tabular} & 1003          & \begin{tabular}[c]{@{}c@{}}16 timeslots (3.2 ms)\end{tabular} & 6971 & 99.29\%
\\ \cline{1-5} 
\end{tabular}
\label{t1}
\end{table}

\begin{figure}
\centering
\includegraphics[scale=0.17]{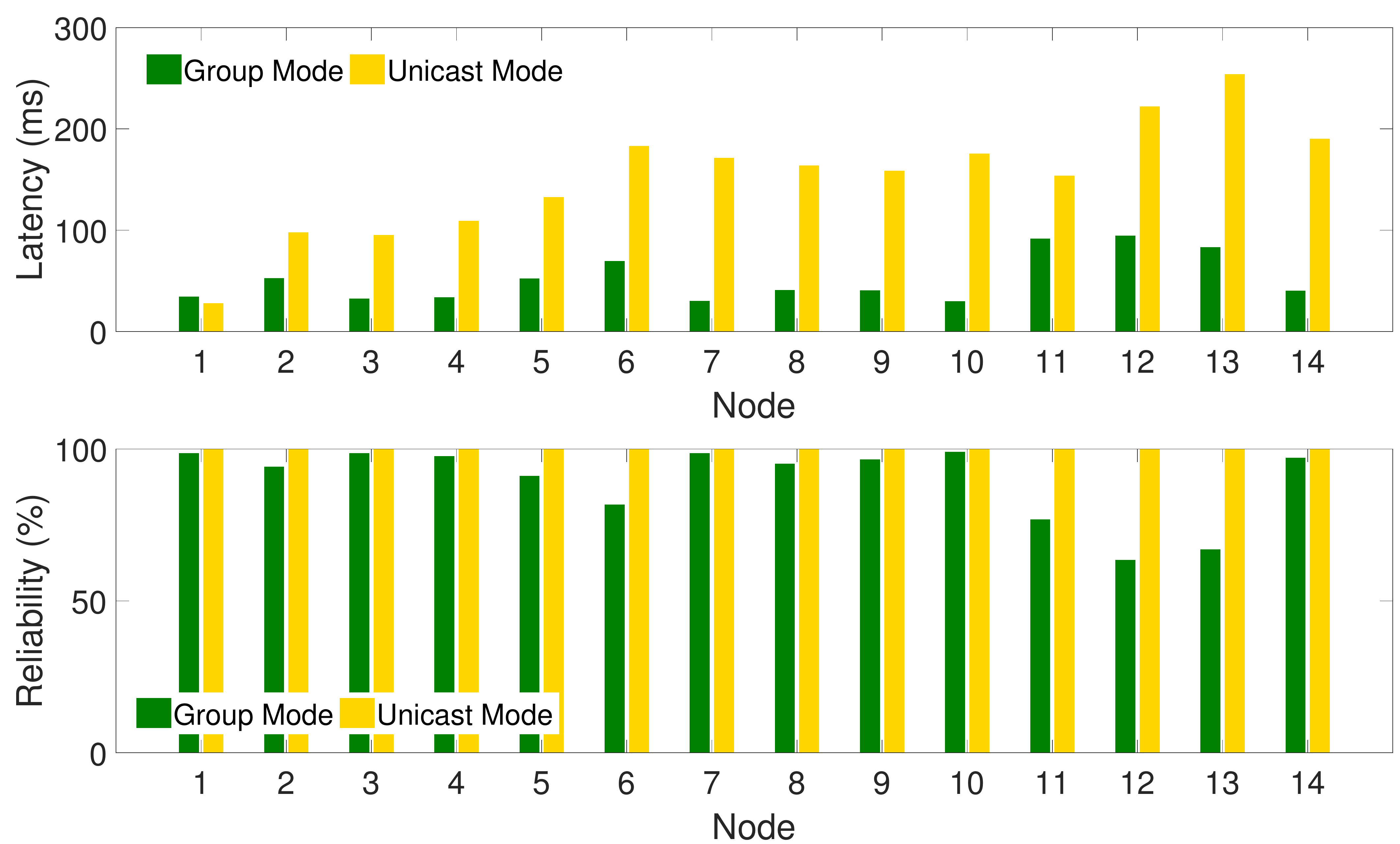}
\caption{Performance evaluation of standard Bluetooth mesh solution.   }
\vspace{-0.5cm}
\label{mesh_res}
\end{figure}

We have conducted a performance comparison of \textsf{GALLOP} against the standard Bluetooth mesh protocol \cite{BT_mesh}.  On the same testbed, we consider  a mesh topology of \(15\) nodes with a maximum of \(3\) hops. Bluetooth mesh protocol is based on a \emph{managed flooding} technique wherein a message injected in the  network is potentially  forwarded by multiple relay nodes. A source node broadcasts data on three distinct channels (known as advertising channels) which are randomly scanned by the relay nodes. 
 We consider a closed-loop control scenario as employed for \textsf{GALLOP}. One of the nodes is the controller which transmits a control message of \(10\) bytes payload to all the slave nodes  in the network. The slave nodes respond by transmitting a response message.   In the \emph{group} mode, the controller floods a group message to all the slave nodes, whereas in the \emph{unicast} mode, the controller sends a control message to each slave node individually. The evaluation results are shown in Fig. \ref{mesh_res}. 
Note that the cyclic exchange is not complete until the controller has received messages from all slaves. Therefore, the worst-case latency is important, which is \(253\) ms and \(94\) ms, in case of unicast mode and group mode, respectively. The unicast mode achieves \(100\%\) reliability as the controller keeps transmitting until it has received acknowledgement from a slave node. For group mode, the controller transmits the message over three 
\emph{advertising events}. It achieves a mean reliability of \(89.6\%\). 
Results show that 
\textsf{GALLOP} outperforms Bluetooth mesh by several orders of magnitude in terms of latency (cycle time) while providing similar or better reliability. 

\subsection{Evaluation on LE Coded PHY}
We have tested the cycle time and reliability performance of \textsf{GALLOP} on the LE Coded 125 kbps PHY as well. The minimum required slot duration for a \(16\) byte PDU is 1.6 ms. To account for processing overheads, we have fixed the timeslot length to \(2\) ms. The transmit power is \(9\) dBm. \tablename~\ref{tcoded} summarizes the key results obtained from evaluation on  Topology 5. 
We have evaluated the reliability of LE Coded PHY in the absence of channel hopping and schedule extrapolation. As shown by the results in \tablename~\ref{tcoded}, the LE Coded PHY incurs significantly higher latency as compared to the LE 2M PHY. In the absence of channel hopping and retransmissions, the achievable reliability is \(96.18\%\). The reliability improves to \(99.97\%\) with schedule duplication. Note that the LE Coded PHY is more vulnerable to interference than the LE 2M PHY due to more time on the air.

\begin{table}[]
\centering
\caption{Summary of Testbed Evaluation (LE Coded PHY)}
\begin{tabular}{ccccc}
\cline{1-5}
				\toprule
Topology                                                    &  Phases & Cycle Time                                                    & \begin{tabular}[c]{@{}c@{}}Successful \\ Packets\end{tabular} & PDR     \\ \cline{1-5} \midrule
\multicolumn{5}{c}{\textbf{No Retransmissions; No Frequency/Channel Hopping}}\\
\midrule
\begin{tabular}[c]{@{}c@{}}Topology 5\end{tabular} & 293403          & \begin{tabular}[c]{@{}c@{}}7 timeslots (14 ms)\end{tabular} & 1410920 & 96.18\%
\\ 
\cline{1-5} 
\midrule
\multicolumn{5}{c}{\textbf{Schedule Duplication without Frequency/Channel Hopping}}\\
\midrule 
\begin{tabular}[c]{@{}c@{}}Topology 5\end{tabular} & 99602          & \begin{tabular}[c]{@{}c@{}}14 timeslots (28 ms)\end{tabular} & 497851 & 99.97\%
\\ \cline{1-5} 
\end{tabular}
\label{tcoded}
\end{table}

\subsection{\textcolor{black}{Evaluation in Dense Deployments}}
\begin{figure}
\centering
\includegraphics[scale=0.6]{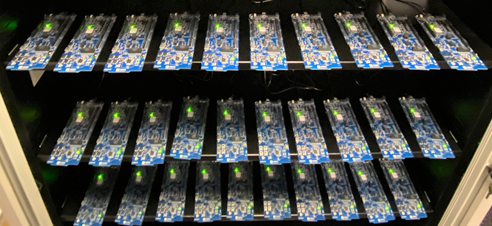}
\caption{\textcolor{black}{Testbed for \textsf{GALLOP} evaluation in dense environments.  }}
\vspace{-0.5cm}
\label{dense_top}
\end{figure}

\textcolor{black}{We have also evaluated the performance of \textsf{GALLOP} in dense deployments. Such deployments represent various control-centric industrial applications including motion control in factory automation (that requires connectivity on a single industrial machine) and wireless battery management systems (monitoring and control of individual battery cells). As shown in Fig. \ref{dense_top}, we have deployed 30 Nordic nRF52840 boards inside a cabinet. This creates a single-hop topology with one controller and \(29\) slaves. We have fixed the timeslot duration for data transmission to \(0.5\) ms to transmit larger payloads while accounting for processing overheads. Our evaluation results, which are based on the LE 2M PHY, are summarized in \tablename~\ref{tdense}. As shown by the results, \textsf{GALLOP} achieved a reliability of \(100\%\) with one schedule duplication. The results also show the viability of the convergence of distributed scheduling in dense deployments.   }

\begin{table}[]
\centering
\caption{\textcolor{black}{Summary of Testbed Evaluation (Dense Deployment)}}
\begin{tabular}{ccccc}
\cline{1-5}
				\toprule
\textcolor{black}{Topology}                                                    &  \textcolor{black}{Phases} & \textcolor{black}{Cycle Time                                                   } & \begin{tabular}[c]{@{}c@{}}\textcolor{black}{Successful} \\ \textcolor{black}{Packets}\end{tabular} & \textcolor{black}{PDR }    \\ \cline{1-5} \midrule
\multicolumn{5}{c}{\textbf{\textcolor{black}{No Retransmissions; No Frequency/Channel Hopping}}}\\
\midrule
\begin{tabular}[c]{@{}c@{}}\textcolor{black}{Single-hop}\end{tabular} & \textcolor{black}{7213}         & \begin{tabular}[c]{@{}c@{}}\textcolor{black}{30 timeslots (15 ms)}\end{tabular} & \textcolor{black}{209173} & \textcolor{black}{99.9\%}
\\ 
\cline{1-5} 
\midrule
\multicolumn{5}{c}{\textbf{\textcolor{black}{Schedule Duplication without Frequency/Channel Hopping}}}\\
\midrule 
\begin{tabular}[c]{@{}c@{}}\textcolor{black}{Single-hop}\end{tabular} & \textcolor{black}{7229}         & \begin{tabular}[c]{@{}c@{}}\textcolor{black}{60 timeslots (30 ms})\end{tabular} & \textcolor{black}{209641} & \textcolor{black}{100\%}
\\ \cline{1-5} 
\end{tabular}
\label{tdense}
\end{table}

\subsection{Technology Demonstrations}
We have demonstrated the basic concept of \textsf{GALLOP} and its viability for handling versatile closed-loop control applications with fast dynamics. In \cite{demo_infocom}, we have demonstrated \textsf{GALLOP} for remote control and formation control of mobile platforms. In \cite{demo_ewsn}, we have demonstrated the scenario of remotely balancing a two-wheeled robot over \textsf{GALLOP} that represents an inverted pendulum on wheels. 

\section{Concluding Remarks} \label{sect_cr}
Realizing wireless closed-loop control is crucial for mission-critical IIoT systems. This paper presented \textsf{GALLOP}, which has been designed to provide high-performance connectivity for achieving wireless closed-loop control over multi-hop networks.  \textsf{GALLOP} implements novel control-aware multi-hop scheduling  for handling bi-directional cyclic information exchange and retransmission techniques that leverage cooperative transmissions for successful operation in harsh wireless environments. Performance evaluation through hardware implementation on a Bluetooth 5 platform and extensive system-level simulations demonstrate that  \textsf{GALLOP} provides highly scalable connectivity with very low and deterministic latency and very high reliability.  The results also demonstrate the viability of handling control loops with fast dynamics (on the order of few ms), rapid convergence and single-hop operation.  \textsf{GALLOP} can be extended to a broad range of legacy and emerging IIoT applications.   

\textcolor{black}{\textsf{GALLOP} is one of the first protocols designed for wireless closed-loop control. There are a number of avenues for further work in terms of protocol enhancements, implementation and application design. From protocol perspective, some of the key directions for future work include MAC layer reconfigurability for dynamic adaptation of scheduling parameters, operation of multiple single-hop networks, enhancements for wireless closed-loop control with node mobility, and integration of antenna diversity techniques. From implementation perspective, it is important to realize \textsf{GALLOP} over a Wi-Fi radio for achieving high-speed and high-performance connectivity for multimedia-centric industrial control applications. The high-performance connectivity offered by \textsf{GALLOP} opens up various new possibilities for \emph{communication-oriented} design of robotics-centric applications (e.g., formation control of mobile platforms), as opposed to conventional control-oriented designs which leverage complex sensory capabilities and architectures. Therefore, an important area of future work is exploitation of \textsf{GALLOP} in cooperative robotics applications, especially from the perspective of control-wireless co-design.  }








\bibliographystyle{IEEEtran}

\bibliography{IEEEabrv,GALLOP_bib}
%


\begin{IEEEbiography}[{\includegraphics[width=1in,height=1.25in,clip,keepaspectratio]{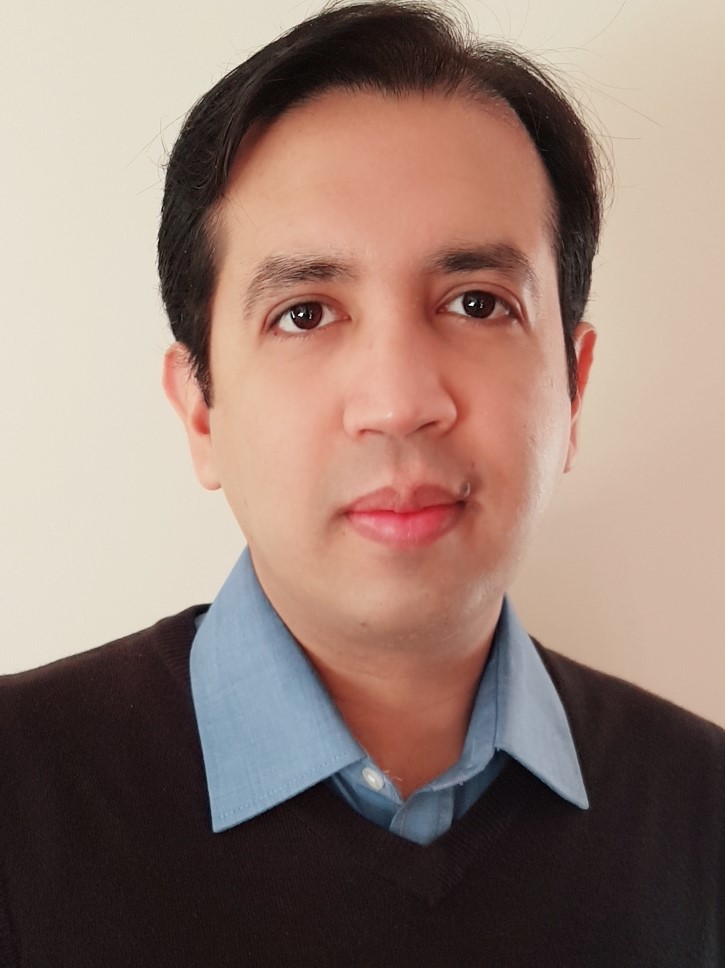}}]{Adnan Aijaz}
(M'14--SM'18) studied telecommunications engineering at King’s College London, United Kingdom, where he received a Ph.D. degree in 2014 for research in wireless networks. He joined the Bristol Research and Innovation Laboratory (formerly Telecommunications Research Laboratory)  of Toshiba Corporation in 2015, where he currently holds the position of Innovation Programme Lead. His recent research areas include industrial communication systems and automation networks, cyber-physical systems, next generation Wi-Fi and cellular (5G and beyond) technologies, and robotics and autonomous systems. He has been contributing to various national and international research projects and standardization activities related to industrial communication.
\end{IEEEbiography}

\begin{IEEEbiography}[{\includegraphics[width=1in,height=1.25in,clip,keepaspectratio]{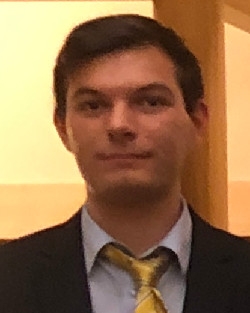}}]{Aleksandar Stanoev} received the M.Eng. degree (Hons.) from the University of Bristol, United Kingdom, in 2018. He is currently a Senior Research Engineer with the Bristol Research and Innovation Laboratory of Toshiba Corporation. His current research interests include low-power wireless MAC protocols, large-scale smart city deployments and their design challenges, and embedded IoT hardware design. He is also a Maintainer of the Contiki-NG open-source IoT operating system.
\end{IEEEbiography}

\end{document}